# Exo 2: Growing a Scheduling Language

| Yuka Ikarashi<br>MIT CSAIL<br>Cambridge, USA | Kevin Qian<br>MIT CSAIL<br>Cambridge, USA | Samir Droubi<br>MIT CSAIL<br>Cambridge, USA |
|---|---|---|
| Alex Reinking<br>Adobe<br>Cambridge, USA | Gilbert Louis Bernstein<br>University of Washington<br>Seattle, USA | Jonathan Ragan-Kelley<br>MIT CSAIL<br>Cambridge, USA |

## Abstract

User-schedulable languages (USLs) help programmers productively optimize programs by providing safe means of transforming them. Current USLs are designed to give programmers *exactly* the control they want, while automating all other concerns. However, there is no universal answer for what performance-conscious programmers want to control, how they want to control it, and what they want to automate, even in relatively narrow domains. We claim that USLs should, instead, be designed to grow. We present Exo 2, a scheduling language that enables users to define *new scheduling operations* externally to the compiler. By composing a set of trusted, fine-grained primitives, users can safely write their own scheduling library to build up desired automation. We identify *actions* (ways of modifying code), *inspection* (ways of interrogating code), and *references* (ways of pointing to code) as essential for any user-extensible USL. We fuse these ideas into a new mechanism called *Cursors* that enables the creation of scheduling libraries in user code. We demonstrate libraries that amortize scheduling effort across more than 80 high-performance kernels, reducing total scheduling code by an order of magnitude and delivering performance competitive with state-of-the-art implementations on three different platforms.

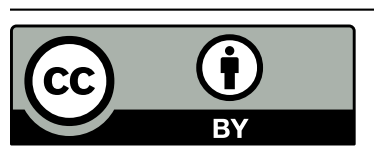

## 1 Introduction

The process of optimizing a program consists of repeatedly modifying it into new programs which compute the *same result* more efficiently. User-schedulable languages (USLs) such as Halide, TVM, TACO, and Taichi offer a promising solution to increase programmer productivity when doing such optimization [10, 28, 39, 54, 56]. USLs reify this optimization process into an explicit *scheduling meta-program* that transforms an underlying *object program* into a better-performing one. By providing formal or informal guarantees that scheduling transformations preserve the equivalence of the object program, USLs enable performance engineers to focus on optimization strategies rather than painstakingly ensuring correctness [57].

Some non-user schedulable languages, such as C/C++ and Python, are designed to "grow" to large projects through the development and composition of libraries built on top of them [63]. These libraries add richer interfaces and functionality to the language without needing modifications to the compiler. Current USLs are designed with a fixed set of scheduling operations that target a particular application domain (e.g., image processing) and/or class of optimization (e.g., loop transformations). Similar to how scripting languages like bash, sed, and awk excel at writing small scripts, USLs excel at optimizing small, individual kernels when the task aligns with the language's capabilities. However, similar to specialized scripting languages, existing USLs are not designed to grow by implementing libraries.

The goal of these fixed sets of scheduling operations is to provide explicit *control* over key optimization choices while abstracting away tedious details. Well-designed USLs must therefore carefully choose a delineation between *automated* optimizations and *user-scheduled* choices exposed as scheduling operations. For instance, the Halide language automates bounds inference, register allocation, and instruction selection (to name a few), but gives users explicit control over loop tiling, fusion, and work vs. locality tradeoffs. When the design works, it shields performance engineers from unnecessary concerns, improving productivity.

In practice, however, it is difficult to design this automation-control boundary perfectly. When the design breaks down, performance engineers have no way to cross the boundary and must drop down to C or assembly code (or even modify the USL compiler) to regain control. These failures can happen either because an optimization is not expressible with the provided controls, or because of imperfect automation. For example, prefetching was initially automatic in Halide but later became user-schedulable, and fixed-point vector instruction selection still requires substantial research and engineering efforts to improve its automation, more than a decade after Halide's initial release [2, 40, 58]. Targeting novel hardware accelerators introduces additional challenges. Halide has yet to support tensor accelerators (e.g., Gemmini, Tensor Core, TPU) because they pose many non-trivial questions about the separation of control and automation. For example, should Halide automatically handle instruction selection for accelerators as well? If so, to what extent can the existing vector instruction selection logic be reused? If not, how should Halide expose control over instruction selection to users? Such design failures are inevitable and generate pressure on USL implementers both to extend the scheduling language and to improve automation.

Furthermore, even in an ideal world where USLs provide users ample control and automation in their scheduling operations, users would still face challenges when implementing realistic HPC kernels. Leading HPC libraries, such as OpenBLAS, cuDNN, and MKL, provide a wide range of configurations through their APIs. With existing USLs, users must laboriously write different schedules for each of these configurations—a cross-product of operations, data types, operational parameters, storage formats, and target architectures. This approach quickly becomes unwieldy and cannot scale effectively. Thus, providing users with means to abstract and automate schedule generation is crucial to effectively manage this vast configuration space.

Therefore, we believe we should design USLs that allow users to (i) build automation *external* to the compiler and (ii) decide to use more or less automation in different contexts. We propose Exo 2, a scheduling language that empowers users to safely build *new scheduling operations* from a set of *primitive*, equivalence-preserving operations. Exo 2 is a new version of Exo [31] and is publicly available under the MIT license[1]. By composing safe primitives, users can write libraries external to the compiler and automate common optimizations, allowing schedule reuse across different kernels while retaining safety guarantees. When users require more control, they can always fall back to using lower-level primitives, satisfying both aforementioned criteria.

Drawing an analogy from the classic perception-control model in robotics and cognition, we conceptualize scheduling languages through the **actions** they take on object code (Section 3), their **inspection** or perception of that code (Section 4), and the **references** pointing to the object code (Section 5). Exo 2's reference mechanism, *Cursors*, binds these three pillars together and enables the creation of scheduling libraries within user code (Figure 1, middle). With this, we demonstrate libraries which amortize scheduling effort across more than 80 high-performance kernels, including BLAS level 1, 2, GEMM, blur, and unsharp mask (Figure 1, top). These libraries reduce total scheduling code by an order of magnitude and deliver performance comparable to or better than MKL, OpenBLAS, BLIS, and Halide on AVX2, AVX512, and Gemmini platforms.

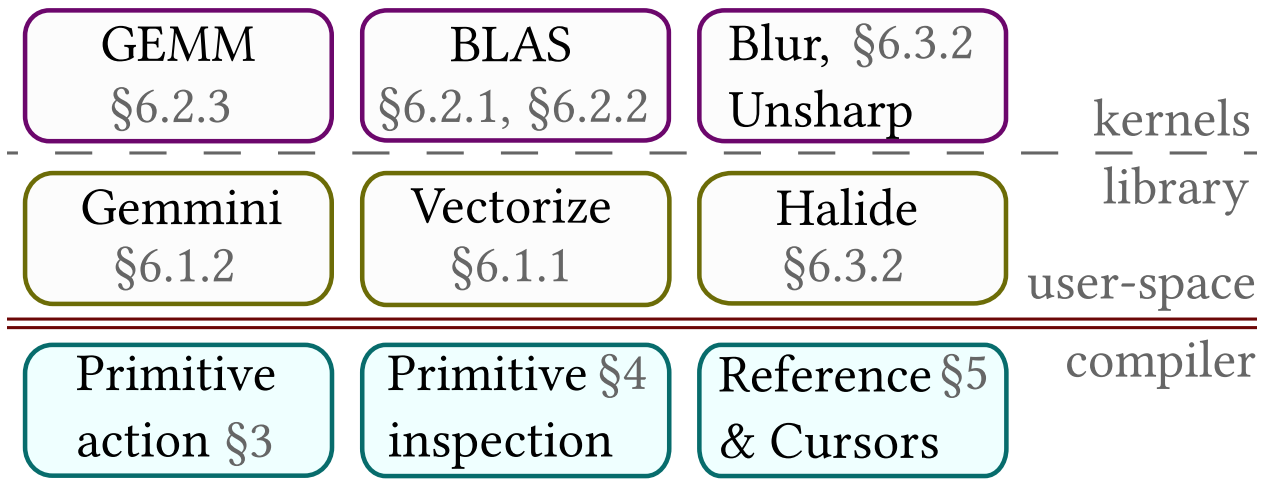

**Figure 1.** Paper overview

The full version of this paper, including appendices with a list of all Exo 2 primitives, library and kernel code for GEMM, and performance graphs for BLAS, can be found in [32].

## 2 Preliminaries

We implemented Exo 2 on top of the Exo *object* language [31]. Exo object code will be framed in boxes to distinguish it from Exo 2 *scheduling* code (without the frame) throughout the paper. We introduce Exo with a matrix-vector multiply example (gemv, inset above), although the specifics of Exo are mostly incidental to this paper.

```
def gemv(M: size, N: size,
         A: f32[M, N] @DRAM,
         x: f32[N] @DRAM,
         y: f32[M] @DRAM):
  assert M % 8 == 0
  assert N % 8 == 0
  for i in seq(0, M):
    for j in seq(0, N):
      y[i] += A[i, j] * x[j]
```

Both procedure arguments and variable declarations follow the syntax ⟨*name*⟩:⟨*type*⟩[⟨*size*⟩] @ ⟨*memory*⟩. In gemv, `f32` is a numeric type for 32-bit floating point. The ⟨*size*⟩s may be constants, or refer dependently to other arguments (e.g., `[M,N]`) as they do here. The `@` symbol, followed by a memory space identifier, specifies the memory space where the variable or argument resides (e.g., `@DRAM`). The `assert`s guarantee that all sizes are even multiples of 8, which can be exploited during scheduling. Finally, `for i in seq(0, M)` is a *sequential* for loop that iterates from 0 to $M - 1$, inclusive.

In Exo 2, schedules are Python meta-programs that take a procedure (e.g., gemv) as input and return a functionally equivalent, rewritten procedure as output. For example, the `divide_loop(p, loop, factor, new_vars, ...)` scheduling operator divides a single loop of $n$ iterations into a pair of outer and inner loops of $n/factor$ and *factor* iterations. It takes the procedure, the loop to be divided, the division factor, new iterator variable names, and optional "tail strategy" arguments for handling cases where $n$ does not divide evenly by *factor*, and returns the modified procedure.

[1] https://github.com/exo-lang/exo

This raises a natural question: how can we specify which loop we want to divide? As we will explore in Section 5, the problem of *reference* is a fundamental issue in scheduling languages. To get us off the ground, we will introduce two mechanisms for identifying object code. First, we can refer to parts of the object code by *name*. For example, we can divide the `'i'` loop of `gemv` simply by naming it: `divide_loop(gemv, 'i', ...)`. Second, we can refer to more complex or specific pieces of code structurally using more general *patterns*. For example, we could refer to that same `'i'` loop in `gemv` with the pattern `'for i in _: _'`. This pattern matches a loop with the iteration variable `i`. The wildcards (`_`) match any range specification and any loop body. Finally, we reify these references in objects we call *Cursors*, drawing an analogy to cursors in text editors. Procedures expose `find_loop` and `find` methods that take loop names or patterns and return matching cursors:

```
cur_0 = gemv.find_loop('i')          # loop name
cur_1 = gemv.find('for i in _:_')    # pattern
assert(cur_0 == cur_1)  # points to the same loop
```

Most scheduling operators take cursors for any reference arguments, though they often also accept pattern strings as an optional convenience. For example, `divide_loop(p, 'i', ...)` is shorthand for `divide_loop(p, p.find_loop('i'), ...)`.

## 3 Action

The core of any USL is a set of scheduling operations tailored to specific applications, optimization classes, and target programmers. In order to design a USL for growth, the set of scheduling operations should enable users to (i) develop libraries for automation while (ii) retaining the ability to resort to low-level control when necessary. While many USLs' scheduling operation sets are designed to fit specific use cases, Exo 2's design revolves around composing low-level primitives. Instead of building in scheduling operations closely matched to what we think users will want, we argue that scheduling languages should focus on providing the essential composable building blocks to ultimately be able to achieve those same transformations. We call these *scheduling primitives*. Primitives should be fine-grained, offering control over low-level details and enable a wide range of transformations through composition. Although composition can always provide abstraction, it cannot provide *finer-grained* control than the underlying primitives themselves. Primitives should also be *safe*: the language implementation should verify that the program remains functionally equivalent after applying a primitive. `divide_loop` and `lift_scope` are examples of such primitives used in the following sections. We implemented a rich set of 46 scheduling primitives for Exo 2, which is shown in Appendix A [32]. These low-level primitives provide users with precise control but reduce the degree of automation provided. In the following sections, we show various ways of composing these scheduling primitives to build automation.

### 3.1 Sequential Composition of Primitives

Suppose we want to write a schedule that tiles a loop nest, a common optimization for improving data locality. Schedules can often be expressed as a series of rewrites, and these rewrites can be composed sequentially. Each scheduling operation returns a new procedure that can be further scheduled by the next operation. For example, we can tile `gemv` by sequentially composing `divide_loop` and `lift_scope` primitives as follows. For brevity, we abbreviate `gemv` to just `g`.

```
g = divide_loop(g,'i',8,['io','ii'],perfect=True)
g = divide_loop(g,'j',8,['jo','ji'],perfect=True)
g = lift_scope(g,'jo')
```

```
for io in seq(0, M/8):
  for jo in seq(0, N/8):
    for ii in seq(0, 8):
      for ji in seq(0, 8):
        y[8*io+ii] += A[...]*x[...]
```

Since we know that the factor 8 perfectly divides both `M` and `N`, we can omit code for handling tail cases by passing `perfect=True` to `divide_loop`. `lift_scope` takes either a `for` loop or an `if` statement and interchanges it with the surrounding `for` or `if`. This simple composition of primitives yields the tiled `gemv` object code as shown above.

### 3.2 Reusable Schedule Fragments as Functions

Since tiling is a common optimization, many existing USLs provide built-in tile scheduling operations. However, there is no need to provide tiling as a built-in: function abstraction offers a natural way to encapsulate such scheduling patterns as a reusable, user-level scheduling function.

```
def tile2D(p,                # procedure
           i_lp, j_lp,       # loop names
           i_itrs, j_itrs,   # list of new names
           i_sz, j_sz):      # tile sizes
  p = divide_loop(p,i_lp,i_sz,i_itrs,perfect=True)
  p = divide_loop(p,j_lp,j_sz,j_itrs,perfect=True)
  p = lift_scope(p, j_itrs[0])
  return p
```

Now, instead of calling scheduling primitives directly, we achieve the same transformation with a single call to `tile2D`:

```
g = tile2D(g,'i','j',['io','ii'], ['jo','ji'], 8,8)
```

Functionally, `tile2D` has the same behavior as if it were a built-in scheduling operation. Each Exo 2 primitive has the type Op = Proc × Cursor × ... ⟶ Proc, taking a procedure, a cursor, and other arguments and returning a procedure. This design contrasts with the more common method chaining style used in Halide [55], TVM [10], and the original scheduling language design of Exo [31], where scheduling operators are methods on a program object. Our design enables users to create user-defined scheduling operations with the same type, allowing for seamless integration of primitives and user-defined functions. With the expected type Op, `tile2D` is indistinguishable from a built-in scheduling primitive.

### 3.3 Schedules with Control Flow

Traditional control flow constructs are useful when developing new scheduling operations and automation. Conditionals enable schedules to perform different actions based on the situation, such as applying different schedules depending on the precision or hardware target. Loops allow us to parameterize the notion of repeating an action. For example, the `tile2D` function could be generalized to work on an arbitrary number of dimensions:

```
def tilenD(p, loops, new_iters, tile_sizes):
    for i, loop in enumerate(loops):
        p = divide_loop(p, loop, tile_sizes[i],
                        new_iters[i], perfect=True)
    for i, _ in enumerate(loops):
        for j in range(0, i):
            p = lift_scope(p, new_iters[i][0])
    return p
```

Since Exo 2 scheduling primitives are guaranteed to raise an error when the transformation breaks functional equivalence, exception handling mechanisms like `try/except` can thus implement different behaviors based on the safety of scheduling primitives. This approach generalizes `tile2D` to non-perfectly divisible loops as shown below.

```
def general_tile2D(...): # same signature as tile2D
  orig_p = p
  try:
    p = tile2D(p, ...) # call to tile2D by default
  except:
    p = divide_loop(orig_p, ..., tail="guard")
    p = divide_loop(p, ..., tail="guard")
    p = lift_scope(p, new_j_iters[0])
    p = lift_scope(p, new_j_iters[0])
  return p
```

This schedule first tries to perfectly tile the code by calling `tile2D`, and if unsafe (i.e. it raises an exception), it defaults to a general tiling schedule by re-starting the process using the tail strategy (`tail="guard"`) which adds a guard to avoid out-of-bounds accesses. Since two calls to `divide_loop` generate two `if` guards, we need to lift the outer loop (`new_j_iters[0]`) twice to achieve the desired tiled loop ordering.

In practice, there are three distinct types of user-facing errors that can occur, and it is beneficial to differentiate between them. The first type of error is the *SchedulingError*, which is raised by the compiler analysis when user code attempts to apply transformations that do not preserve functional equivalence. The second type is the *InvalidCursorError*, which is triggered when navigating to invalid locations (see Section 5.2). Lastly, there may be internal compiler errors caused by compiler implementation bugs or other internal issues. It is preferable to specify the type of error being caught, for example, by using `except SchedulingError`, rather than catching all exceptions indiscriminately.

### 3.4 Higher-order Scheduling Functions

We define operations of type $\widehat{\text{Op}} = \text{Proc} \times \text{Cursor} \times \ldots \rightarrow \text{Proc} \times \text{Cursor}$. Any Op, defined in Section 3.2, can be lifted to an $\widehat{\text{Op}}$ by lift $op = \lambda(p, c).(op(p), c)$. This allows us to define higher-order scheduling combinators that take $\widehat{\text{Op}}$s as arguments and produce a $\widehat{\text{Op}}$ as output.

```
def seq(*ops):
  def func(p, c, *args):
    for op in ops:
      p,c = op(p,c,*args)
    return p, c
  return func
```

```
def repeat(op):
  def func(p, c, *args):
    try:
      while True:
        p,c = op(p,c,*args)
    except:
      return p, c
  return func
```

```
def try_else(op, opelse):
  def func(p,c,*args):
    try:
      p,c = op(p,c,*args)
    except:
      p,c = opelse(p,c,*args)
    return p, c
  return func
```

```
def reduce(op, top):
  def func(p,cur,*args):
    for c in top(cur):
      p,c = op(p,c,*args)
    return p, c
  return func
```

Higher-order scheduling functions are implemented using these combinators. For example, `seq(lift_alloc, lift_alloc)` is a scheduling function that lifts an allocation twice, while `repeat(lift_alloc)` lifts an allocation as much as possible. Both functions retain the type $\widehat{\text{Op}}$, and so can be further combined as expected.

## 4 Inspection

Type reflection, also known as introspection for homogeneous systems and inspection for heterogeneous systems, is a meta-programming feature that enables programs to dynamically examine object code through built-in functions (e.g., `typeOf` in Haskell) or pattern matching. To the best of our knowledge, no existing USL exposes inspection of object code to users, making operations such as "unroll all loops with bounds less than 5" inexpressible. We believe that inspection is essential for effective meta-programming in user-extensible USLs. For example, when creating a vectorizer library, identifying properties such as reductions, buffer precisions, and loop-invariant vectors in the object code is crucial for determining parallelization strategies and choosing vector instructions. Figure 4 and Section 6.1.1 demonstrate how the presence of FMA instructions affects optimal staging, necessitating inspection to identify them.

Exo 2 provides type reflection through cursors, allowing users to examine standard AST properties like variable names, literal expression values, and annotations (e.g., memory spaces and precisions) at scheduling time. Cursors also support local AST navigation, accessing loop bounds (`loop.hi()`) and bodies (`loop.body()`). More complex inspection operations can be built by combining inspection and navigation primitives. For instance, bounds inference, which determines all possible index accesses to an array within a given scope, is provided as a *built-in* feature in Halide. In contrast, Exo 2 provides users with facilities to implement bounds inference as a new inspection operator *external* to the language, which we can then use as a building block of the Exo 2 Halide library (Section 6.3.2). As an example, consider inferring access bounds for `arr` within the `io` loop:

```
for io in seq(0, N / 32):
    # arr is accessed within [32 * io : 32 * io + 34]
    for ii in seq(0, 32):
        x = arr[32 * io + ii] + arr[32 * io + ii + 1]
                              + arr[32 * io + ii + 2]
```

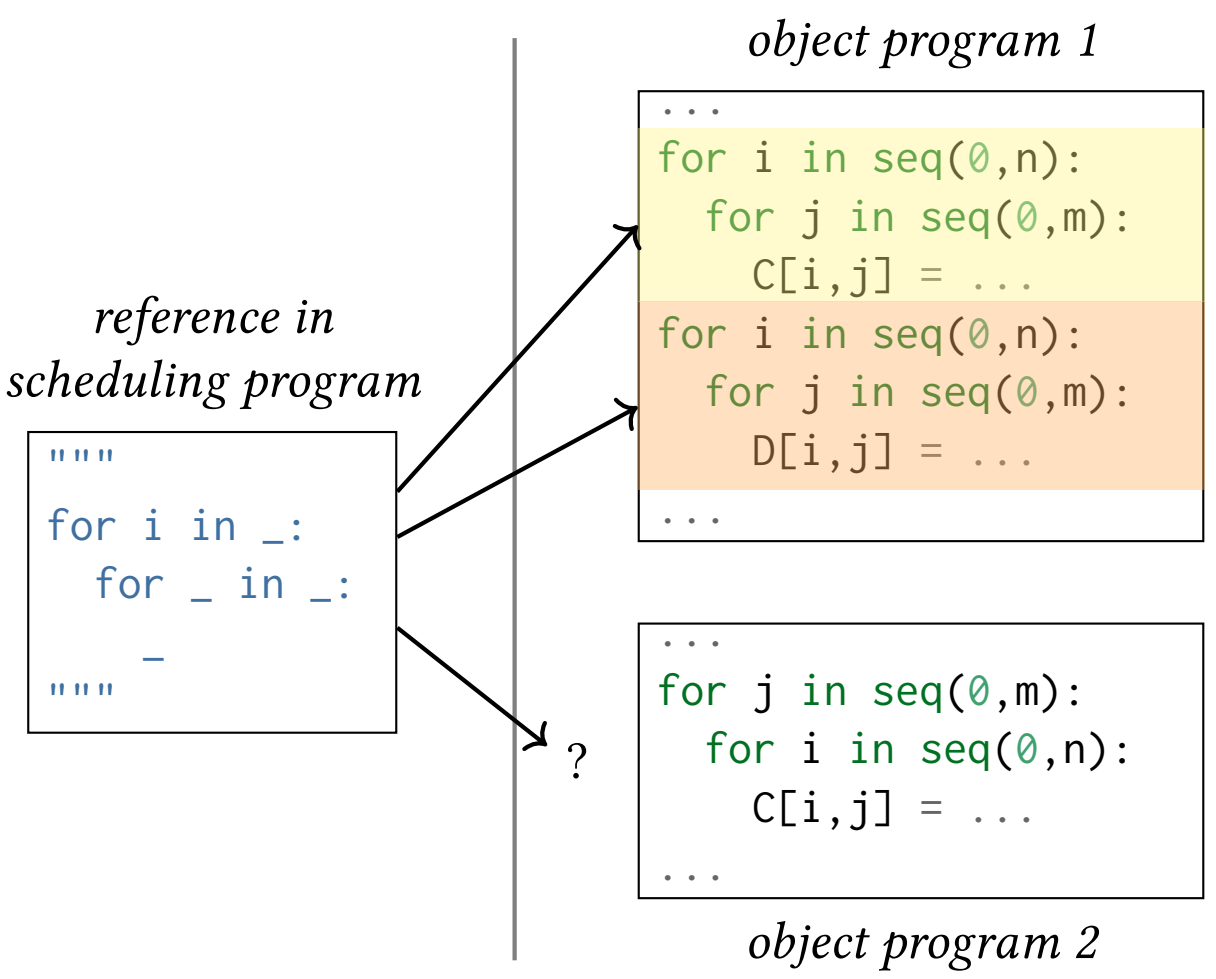

**Figure 2.** Schedules need mechanisms to *refer* to the object code they wish to modify. The meaning of a given reference expression is often dependent on context (*frame of reference*), and may refer to zero, one, or many parts of the program.

Determining the bounds for the array `arr` can be reduced to unioning bounds for each array access. Each index expression is an affine expression consisting of constants and variables which may be either free or bound in the scope. In the index expression `32 * io + ii + 1`, `ii` is a bound variable and `io` is a free variable. Knowing that $0 \leq$ `ii` $< 32$, the index expression spans the window `[32 * io + 1 : 32 * io + 33]`. Our implementation combines primitive cursor inspections, which query the loop bounds expressions, with ordinary Python code which maintains the environment of free/bound variables and unions the inspected bounds.

# 5 Reference

## 5.1 References in USLs

***Nominal v.s. relative reference.*** Suppose we want to specify the loop nest to which we want to apply `tile2D`. Given a pattern (Figure 2, left), there is no guarantee that this reference is unique (Figure 2, program 1), or that we are referring to anything at all (Figure 2, program 2), because resolving references depends on the context, or *frame of reference*. Therefore, USLs must address questions of ambiguity and invalidity for object program references. Halide, for example, opts for a *nominal* reference system that eliminates ambiguity by design. In Halide, each statement is uniquely identified by the buffer it writes to (e.g., `blur_x`; see Figure 11 for sample code), and each loop within loop nest has a unique iterator variable (`x`, `y`, and `xi`). This allows for globally unambiguous reference of loops using the pair (`blur_x`, `y`).

Global nominal references prevent ambiguity but introduce significant limitations, requiring users to manage *globally distinct* names for each object code fragment. Without context-dependent references, it is impossible to specify transformations such as "parallelize the second most outer loop" or "unroll three stages back". Instead, schedules would be hard-coded with nominal references for specific object code. We propose that *relative* references that depend on the frame of reference are essential for building growable scheduling languages. The "nominal" and "relative" reference distinction is analogous to "proper nouns" and "noun phrases" – proper nouns must refer by name (e.g., Boston, NY), while noun phrases enable context-based references (e.g., cities on the east coast), potentially matching zero, one, or many objects. If we change the context from the US to Canada, relative references remain applicable (though they refer to different objects), while nominal references do not.

***Single v.s. multiple references.*** In contrast to Halide, ELEVATE allows users to systematically define traversal strategies to disambiguate references with respect to a tree traversal order. For instance, `topDown` applies a rewrite at the *first* successful application, while `tryAll` applies the rewrite wherever possible. However, ELEVATE only allows having a single reference at a time. This single-frame limitation in ELEVATE can be inflexible and inconvenient because users always have to navigate relative to a single point of reference. Halide, on the other hand, allows multiple references to coexist. This means that users can refer to multiple specific points in the code simultaneously, using their unique names.

***Stable v.s. one-time references.*** Scheduling is a programming process that evolves over time, meaning that frames of reference have both temporal and spatial dimensions. Suppose there are two references to the object code: one pointing to the loop with iterator `i` (Figure 3 light blue) and the other to the loop with iterator `j` (Figure 3 orange). If we tile loops `i` and `j`, what happens to the second reference? Has the object it's pointing to changed, or has it ceased to exist?

```
for i in seq(0,n):
  for j in seq(0,n):
    C[i,j] = ...
for i in seq(0,n):
  for j in seq(0,n):
    D[i,j] = ...
```

Now consider the same schedule applied to a different program (inset left). In this case, there is no complication because the loop nest to which the second cursor points is unaffected by tiling the first.

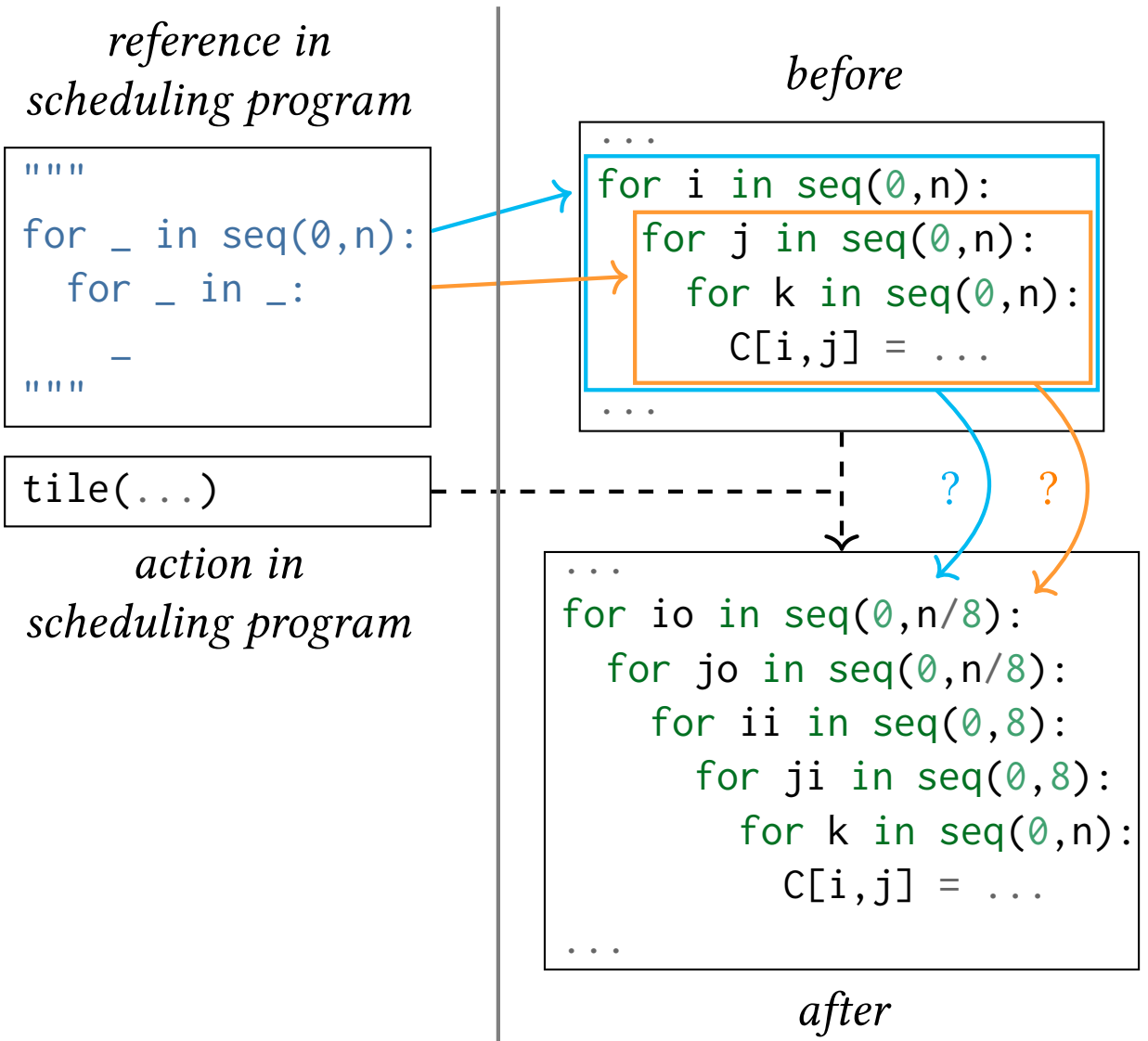

**Figure 3.** The process of transforming code raises the question: where should references to the original code point after a transformation is applied, or should they be invalidated?

These examples demonstrate how references are always implicitly made at a specific moment in the scheduling process. When we resolve a reference, we are essentially asking, "What does this reference denote, *right now*?" To answer this, we identify another taxonomy for USLs: *stable* references that can be reused after applying an action, and *one-time* references that get invalidated after an action. Nominal references, like those in Halide, are effective at providing stable references because they are globally unambiguous. In contrast, ELEVATE's references are one-time, which is similar to other transformative meta-programming systems like jQuery. Pattern matching, as used in these systems, is a brittle mechanism for the scheduling process. A pattern that existed at one point in the scheduling process might not exist or may point to completely different object code after applying an action (e.g., how the `for i` loop ceases to exist after tiling in Figure 3). Furthermore, pattern matching requires exact knowledge of the code structure and its potential transformations, hindering the encapsulation of scheduling operations.

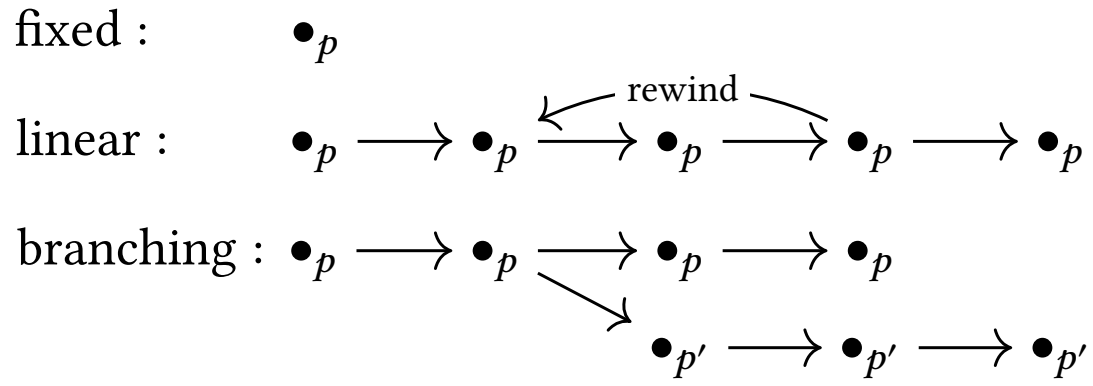

***Scheduling time models in USLs.*** The choice of aforementioned reference designs affects the fundamental design of USLs, and consequently affects the scheduling time models for USLs: (i) The *fixed* time model (inset top) uses a single, static reference frame, largely achieved by Halide's *multiple, stable, nominal* references. (ii) The *linear* time model (inset middle) advances the reference frame for each action and allows it to "rewind" after encountering an error. ELEVATE's *single, one-time, relative* reference supports this model. (iii) The *branching* time model (inset bottom), adopted by Exo 2's *multiple, stable, relative references*, treats time as branching into parallel versions of the procedure, with cursors living at specific versions and scheduling creating different cursors for each branch. The branching time model encompasses the other time models, as we will demonstrate by recreating Halide and ELEVATE-style references in Section 6.3.

## 5.2 Cursors

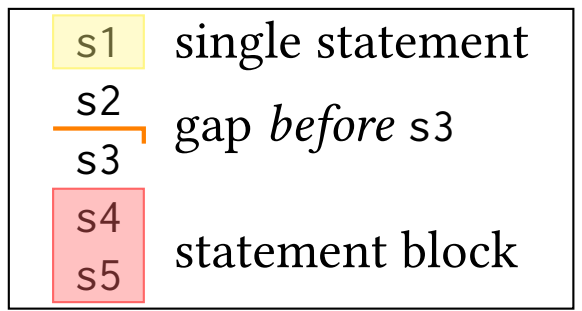

Exo 2's reference mechanism, embodied by *Cursors*, maintains *multiple, stable, relative* references. Cursors, inspired by the blinking vertical bar in text editors, allow users to select and refer to parts of the code such as expressions, blocks of statements, and gaps between statements (inset above).

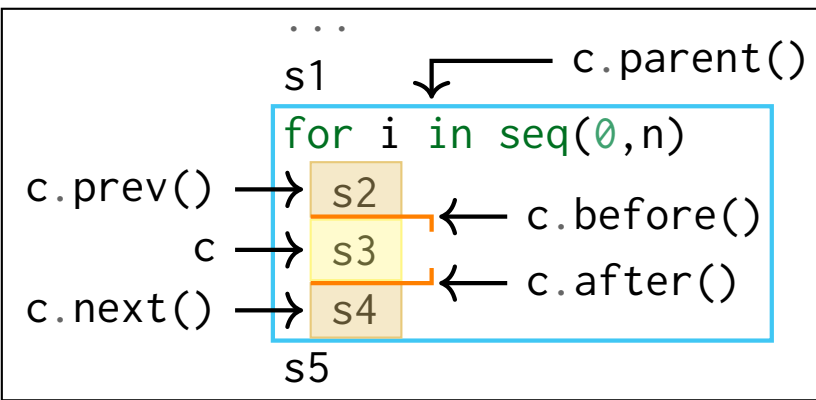

Multiple, relative cursors require modulating the spatial reference frame (navigation) and querying cursors in context. Cursors enable spatial navigation within a procedure (i.e. "after _," "before _," "parent of _," etc.) to proximate locations (see inset above). Moreover, instead of invoking `proc.find(...)`, we can invoke `cursor.find(...)` to restrict our pattern match to the sub-AST identified by the `cursor`. *InvalidCursorError* is raised when navigation is invalid. For instance, `c.parent()` is invalid when `c` is already pointing to the top-level statement.

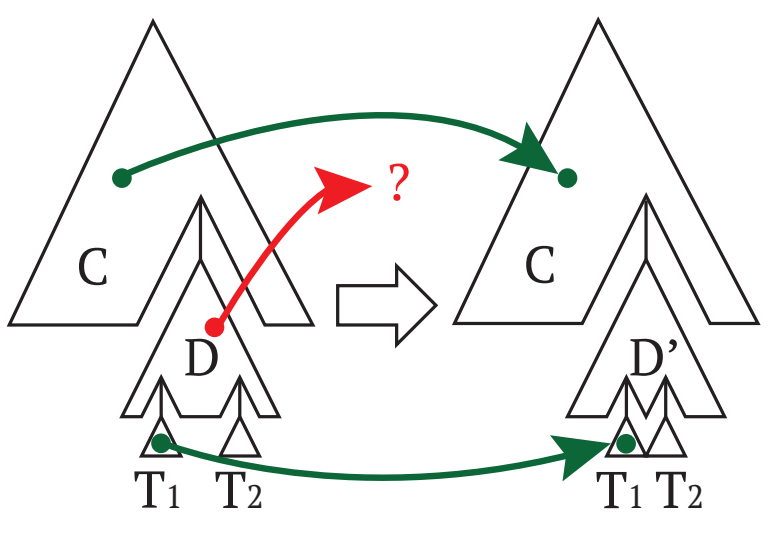

***Forwarding.*** *Cursor forwarding* is the mechanism by which Exo 2 supports stable references. Consider a generic action which rewrites the inset left AST to the inset right AST. We can decompose the AST of the object code as $P = C[S]$, where $C$ is the unmodified subtree and $S$ is the modified subtree. Furthermore, we can usually identify some *invariant* sub-trees $T_1, \ldots, T_k$ of the modified sub-tree $S$ which are not changed by the action. Therefore, we have the further decomposition $P = C[S] = C[D[T_1, \ldots, T_n]]$. Now, we can describe the procedure after the action as $P' = C[D'[T_1, \ldots, T_n]]$ — all changes are isolated to the transformation of $D \rightarrow D'$. Given the decomposition of a program into an outer context $C$, a sub-tree $D$ that is being transformed, and sub-trees $T_1, \ldots, T_n$ that replace $D$, any cursor pointing to a statement in $C$ or $T_i$ can be unambiguously forwarded after the action is completed. Cursors to gaps and statement blocks can be unambiguously forwarded if attached to statements or within $C$ or $T_i$. For composite actions, if a cursor forwarding policy handles the unambiguous primitive action cases correctly, it will correctly forward cursors to any AST part untouched by all scheduling actions.

The ambiguous cases occur when the cursor points to a statement in $D$. There are two high-level design decisions for handling such ambiguous cases: (i) invalidate any ambiguous cursor, which ensures unambiguous behavior but makes scheduling programs more laborious to write, or (ii) attempt to produce a valid cursor whenever possible, which maximizes convenience but may lead to unintuitive behavior. Exo 2 has chosen the second approach and defines heuristic forwarding rules for each scheduling primitive. This can lead to ambiguous forwarding behavior, especially when calling scheduling library functions that compose many forwarding effects. However, since forwarding is deterministic even in ambiguous cases, we find that simply printing cursors after the function call is usually sufficient to understand the behavior in practice.

When a primitive action takes in a procedure and input cursors, Exo 2 will *implicitly* forward the cursors to the procedure's reference frame, since cursors should, by definition, always point to the procedure. Therefore, `expand_dim(p, c, ...)` becomes a shorthand for `expand_dim(p, p.forward(c), ...)`. However, navigation and forwarding do not commute in general. If `c.next()` is passed to a primitive action, the implicit forwarding behavior is equivalent to `p.forward(c.next())`, not `p.forward(c).next()`.

***Implementation.*** Internally, a cursor stores two values: a reference to the procedure it's pointing at (i.e. a *time* coordinate) and a path defining its relative location inside that procedure's AST (a *spatial* coordinate). The path describes navigation in an AST as a downward traversal. In an AST, all children are labeled (e.g. the *rhs* of a binary operation) and are either a node or list of nodes (e.g. the *"body"* of a for loop). Thus, each downward step in the AST may be represented as a label-index pair, where the index is null if the child is not a list.

For example, in this code block, the `y` variable has path $(body, 1)$, $(body, 0)$, $(rhs, \emptyset)$, $(lhs, \emptyset)$. Traverse to the second statement of the procedure body $(body, 1)$, the first statement of the loop body $(body, 0)$, the right-hand-side of the assignment $(rhs, \emptyset)$, and then the left-hand-side of the addition $(lhs, \emptyset)$. Spatial navigation operations such as `.parent()`, `.rhs()`, `.body()[idx]`, or `.next()` are straightforward to implement by modifying this path representation. Cursors to gaps are relative to the statement nodes, and cursors to statement blocks simply store a *range* at the final path index, instead of a single value.

```
x: i8
for j in seq(0, 2):
  x = y + z
```

Whenever a scheduling action is applied to procedure $p$ to produce a new procedure $p'$, the Exo 2 runtime records this provenance along with an internal forwarding function. This internal forwarding function defines how to forward from a cursor $c$ into $p$ to a cursor $c'$ into $p'$. This is done by decomposing the effect of every primitive action into atomic AST edits (insert, delete, replace, move, and wrap; as shown below), each with its canonical forwarding function.

***Insertion.*** Inserting an IR fragment into the AST preserves existing paths. The forwarding function adjusts paths through the insertion point by incrementing pre-existing paths at the appropriate tree level. The orange block cursor illustrates how existing cursors are forwarded when a new `pass` is inserted into a gap.

```
def foo(x: f32 @ DRAM):
gap x = 0.0
    x = 1.0
```

```
def foo(x: f32 @ DRAM):
    x = 0.0
    pass
    x = 1.0
```

***Deletion.*** Deleting a subtree from the AST invalidates paths within the subtree (shown as a violet cursor below), while paths outside remain valid (orange cursor). The forwarding function decrements pre-existing paths past the deletion point at the appropriate tree level.

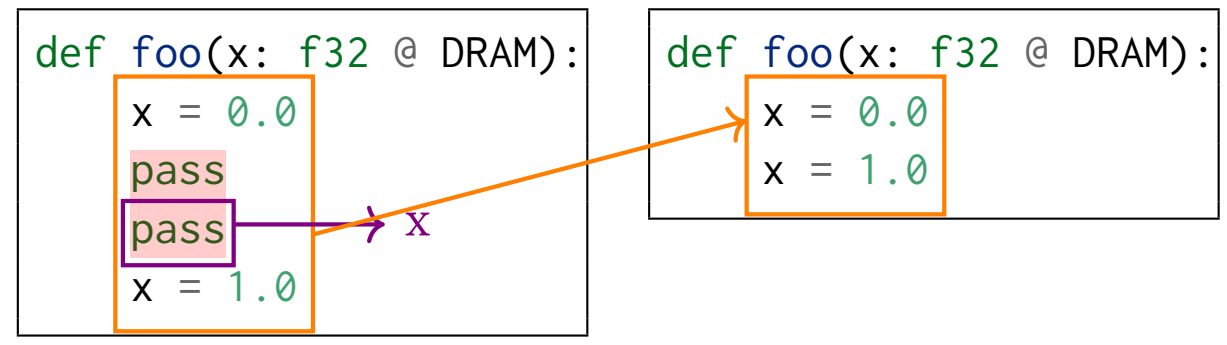

***Replacement.*** Replacement supports higher-level operations like algebraic simplification by replacing an existing subtree (two statements with a red background below) with a new one (`x = 1.0`). The forwarding behavior is similar to inserting the new subtree and deleting the old, except the unique path to the old subtree remains valid.

```
def foo(x: f32 @ DRAM):
    ...
    x = 0.0
    x += 1.0
    ...
```

```
def foo(x: f32 @ DRAM):
    ...
    x = 1.0
    ...
```

***Movement.*** Moving a subtree (orange block cursor) to a gap within the AST preserves its node identity but invalidates cursors across the subtree boundary (violet block cursor). The forwarding function maps paths to their natural correspondences, handling cases of moving to higher or lower levels in the AST.

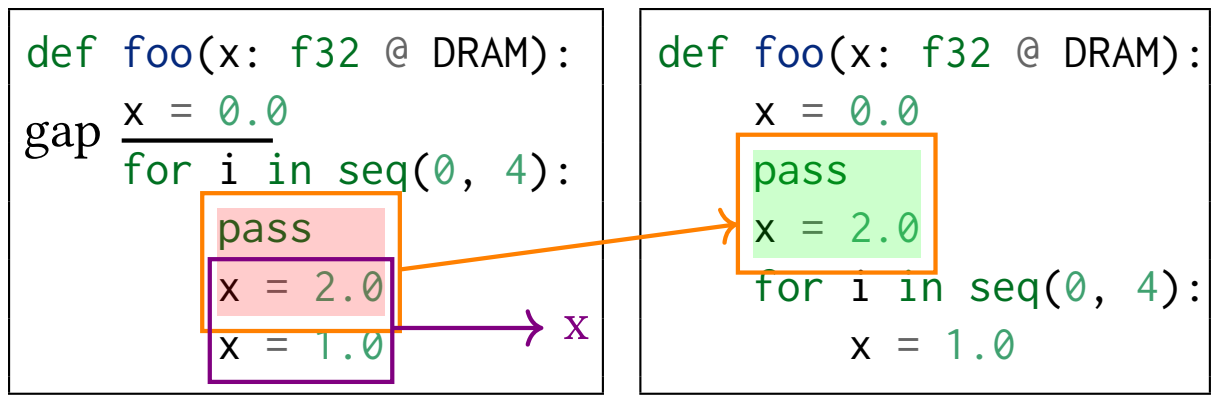

***Wrapping.*** Wrapping an existing subtree (orange cursor) with a one-hole IR fragment (e.g. `for i in seq(0, 4): _`, where the body of the for loop is the hole) is equivalent to insertion with movement for blocks, but not for expressions. Consequently, we separate this function as its own atomic edit. Paths into the wrapped subtree are adjusted by inserting a step at the appropriate level to navigate into the wrapping fragment. Paths through the wrapping point are decremented at the level by the length of the replaced subtree minus one.

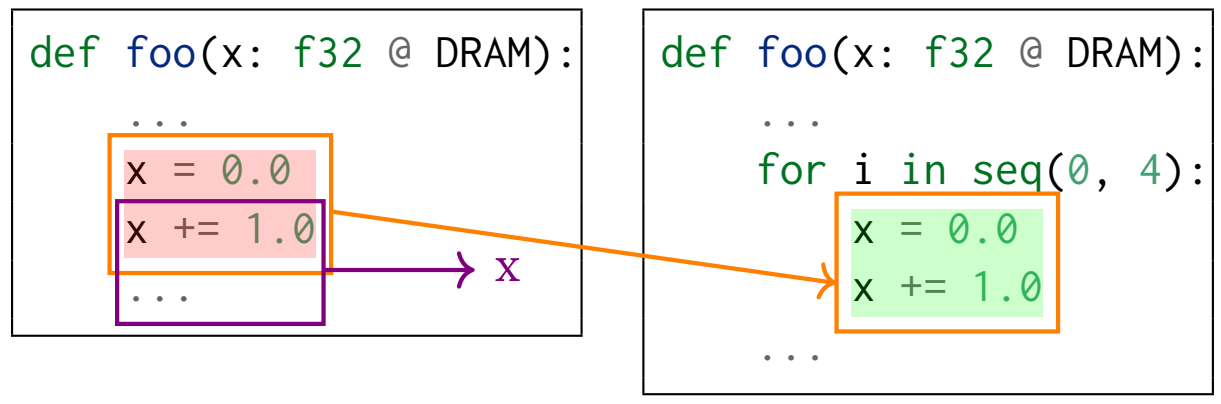

The internal forwarding function for a primitive action is the sequential composition of forwarding through its constituent atomic edits. For example, suppose `p` is the result of applying a primitive action to cursor `c`, and this primitive action was comprised of $n$ atomic edits. When `p.forward(c)` is called, Exo 2 composes the forwarding functions ($f_1$, $f_2$, ..., $f_n$) for each primitive action between `p_0 = c.proc()` and `p_n = p` (the output procedure). This composition produces a single function $f = f_n \circ f_{n-1} \circ \cdots \circ f_1$ that maps `c_0` from its location in the original procedure to the corresponding location in $p_n$. Therefore the forwarded cursor can be obtained by computing $f(c)$. If any forwarding function returns an invalid cursor, invalidity is maintained by each subsequent forwarding function.

## 6 Building Scheduling Libraries

### 6.1 Target-specific Functions

Exo 2's action, inspection, and reference empower users to create scheduling libraries external to the compiler. This section shows how Exo 2 enables users to automate hardware architecture-specific optimization strategies while maintaining control over low-level performance considerations.

**6.1.1 Vector Architectures.** We define a new `vectorize` scheduling operator in user code, which is parameterized over vector width, precision, memory type, and vector instructions, so it can be instantiated for many different vector machines. By leveraging the safety guarantees of the `fission` primitive, which checks for loop-carry dependencies, we eliminate the need for separate dependency analyses, allowing for the concise implementation of transforming loop operations into their SIMD equivalents. Programmers can customize the vectorization process by scheduling deviations as a prologue (e.g., Common Subexpression Elimination) or epilogue (e.g., Loop Invariant Code Motion), or by incorporating hooks to modify behavior in certain cases (e.g., detecting a Fused Multiply-Add, or FMA).

```
for i in seq(0, n):
  y[i] += a * x[i]
```

**(a)** Input saxpy kernel

```
for io in seq(0, n / 8):
  for ii in seq(0, 8):
    var1: f32 @ VEC
    var2: f32 @ VEC
    var2 = y[8*io + ii]
    var3: f32 @ VEC
    var4: f32 @ VEC
    var4 = a
    var5: f32 @ VEC
    var5 = x[8*io + ii]
    var3 = var4 * var5
    var1 = var2 + var3
    y[8*io + ii] = var1
```

**(b)** Staging without FMA

```
for io in seq(0, n / 8):
  for ii in seq(0, 8):
    var1: f32 @ VEC_AVX2
    var2: f32 @ VEC_AVX2
    var2 = y[8*io + ii]
    var3: f32 @ VEC_AVX2

    var3 = a
    var4: f32 @ VEC_AVX2
    var4 = x[8*io + ii]
    var1=var2+var3*var4

    y[8*io + ii] = var1
```

**(c)** Staging with FMA

**Figure 4.** Code transformations for staging the computation (step 3 of `vectorize`), depending on the presence of FMAs.

```
def vectorize(p, loop, vw, precision,
              mem_type, instrs, rules=[]):
    p = divide_loop(p, loop, vw, tail="cut")
    p = parallelize_reductions(p, loop, ...)
    inner = p.forward(loop).body()[0]
    p = stage_compute(p, inner, ..., rules)
    p = fission_into_singles(p, inner)
    return replace_all_stmts(p, instrs)
```

The signature of `vectorize` follows the type $\text{Op} = \text{Proc} \times \text{Cursor} \times \ldots \rightarrow \text{Proc}$ (Section 3.2), which takes a procedure and a cursor pointing to the loop to be vectorized. The implementation follows these steps: (1) Expose parallelism by dividing the loop. (2) Parallelize all reductions in the loop. (3) Get a cursor `inner` to the innermost loop, and stage the computation into temporary assign statements, representing unary (e.g., load) or binary (e.g., addition) operations. Figure 4 shows an example object code of this step, depicting the starting loop in Figure 4a and applying the default staging in Figure 4b. Users can overwrite the default automation when necessary. For instance, Figure 4c shows more efficient staging when an FMA instruction is available. Staging behavior can be customized by `rules`, which scan the expression and specify which sub-expressions in the subtree should be recursively staged. (4) Finally, we fission between the statements to generate a loop-based representation of the SIMD operations, which are replaced with the equivalent hardware target instructions.

```
for io in seq(0, 8):
  for jo in seq(0, 8):
    ...
    for ko in seq(0, 8):
      config_ld(stride(A, 0))
      ld_data(16,16,A,...)
    ...
```

**(a)** The `ld_data` instruction, which reads from the configuration state, is prepended with the `config_ld` instruction to set the stride. The goal in this object code example is to hoist an expensive `config_ld` instruction out of the loops.

```
while True:
  try:
    try:
      while True:
        try:
          p = reorder_stmts(p, ...)
        except:
          raise Exception("...")
    except:
      p = fission(p, ...)
      p = remove_loop(p, ...)
  except:
    break
```

**(b)** Schedule for hoisting a single statement to the top of an object program.

```
repeat(
  try_else(
    seq(
      fission_after,
      remove_parent_loop
    ),
    reorder_before
  )
)(p, c)
```

**(c)** Alternatively, the same schedule as in Figure 5b can be implemented using higher order functions defined in Section 3.4. Section 6.3.1 defines the scheduling operations combined with relative references.

**Figure 5.** (a) Gemmini object code and (b,c) two possible schedules for hoisting a Gemmini configuration instruction.

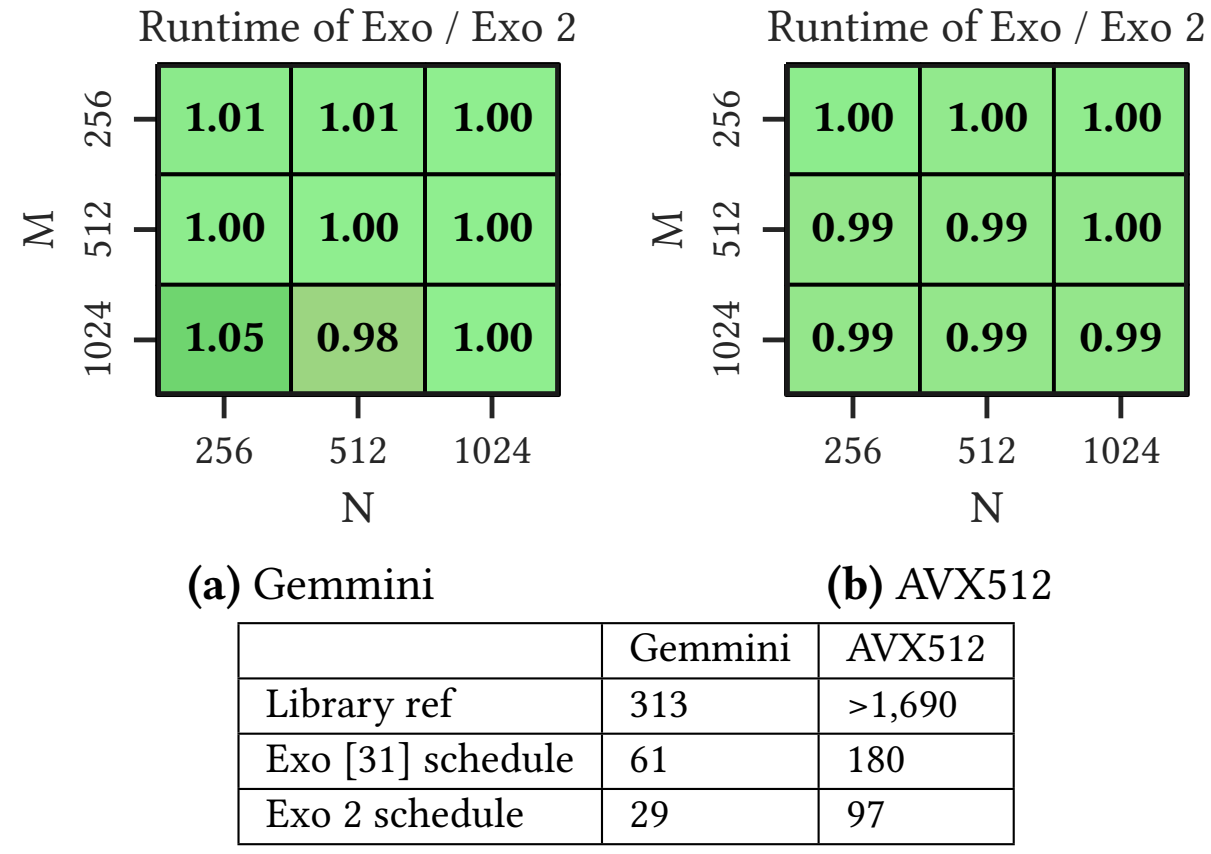

| | Gemmini | AVX512 |
|---|---|---|
| Library ref | 313 | >1,690 |
| Exo [31] schedule | 61 | 180 |
| Exo 2 schedule | 29 | 97 |

**(c)** Lines of code comparison. Exo 2 and Exo schedules are compared against state-of-the-art reference implementations (Gemmini standard library and OpenBLAS).

**Figure 6.** Performance evaluation of Exo 2 against Exo on matmul (higher is better). K is set to 512.

**6.1.2 Gemmini.** Gemmini is a machine learning accelerator developed at UC Berkeley that supports matrix-matrix multiplication (tensor) operations [21]. The GEMM optimization strategy for tensor operations inherently depends on the specificity of the accelerator, and Exo 2 allows users to implement such accelerator-specific optimization passes in a library, which is fully explained in Appendix B [32]. This section focuses on one of the Gemmini library functions, configuration hoisting, which is a unique and specific requirement of stateful accelerators.

Gemmini programmers must program configuration registers to influence the behavior of instructions, such as stride, activation, and quantization. The simplest approach for a compiler is to naively reset this configuration state before every operation, resulting in patterns like `config_ld` followed by `ld_data` (Figure 5a). However, these configuration instructions must be hoisted outside the inner loops to avoid expensive, redundant reissuing. Figures 5b and 5c show two schedules that hoist a single statement as much as possible: one using Python's loop and exception constructs, and the other using higher-order scheduling operations defined in Section 3.4. The statement hoisting schedule repeatedly attempts to reorder the statement to the beginning of the loop, fission the loop, and remove the enclosing loop. This schedule is then used to implement a function that hoists all Gemmini configuration statements.

We implemented matrix-matrix multiply (matmul, gemm) using Exo 2 and compared its performance with Exo. As shown in Figure 6, our implementation achieved similar or faster runtime compared to Exo, which itself is already 3.5x faster than the original Gemmini library [31], while achieving less lines of code than Exo. Performance was measured on the Chipyard version 1.9.1 running Firesim simulator on FPGA, using Gemmini version 0.7.1 [3, 37].

### 6.2 Linear Algebra Library

Unlike existing USLs, Exo 2 allows users to encapsulate application-specific shared scheduling strategies in libraries. This greatly simplifies the optimization of HPC kernels over various configurations—a cross-product of operations, data types, operational parameters, storage formats, and target architectures. We built a linear algebra scheduling library and used it to optimize most of the kernels in BLAS levels 1 and 2, as well as GEMM.

**6.2.1 BLAS Level 1 Operations.** $O(n)$ BLAS operations can be implemented using a single loop. The optimization process involves two main strategies. First, SIMD vector parallelism can be exploited by interleaving loop iterations (modulo the vector width) and autovectorizing the resulting code. Second, the amount of work can be reduced through common subexpression elimination (CSE) and loop-invariant code motion (LICM). Increased register pressure is generally not a concern due to the low arithmetic intensity of level 1 operations. For reductions (dot, asum), computing partial sums in each vector lane exposes data parallelism for vectorization. The degree of loop interleaving is tuned to balance performance gains with register pressure. In practice, CSE and LICM do not typically cause register spilling issues or limit the amount of beneficial loop interleaving that can be performed for BLAS level 1 operations.

We developed a scheduling operator (`optimize_level_1`) that optimizes the entire BLAS level 1 kernels, including asum, axpy, dot, rot, rotm, scale, swap, copy, and dsdot, for a total of 24 kernel variants. However, due to limitations in Exo's object code, which lacks support for value-dependent control, we were unable to support nrm2 and iamax kernels. Appendix D.1 [32] offers the implementation of `optimize_level_1` and performance graphs comparing our results against OpenBLAS, MKL, and BLIS using AVX2 and AVX512 instructions. All results presented in this paper were collected on an AWS m7i.xlarge instance equipped with an Intel(R) Xeon(R) Platinum 8488C processor running at 3.2GHz. Our optimized kernels match the performance of the aforementioned libraries across the board.

#### 6.2.2 BLAS Level 2 Operations.

We show how optimization for $O(n^2)$ BLAS level 2 operations can be shared across different kernels, precisions, and operational parameters.

***General Matrix.*** The common optimization approach for general matrices is turning the loop into a batched program of multiple dot products, allowing vector reuse and reducing memory loads (unroll-and-jam). In contrast to normal loop unrolling, this transformation will jam $c$ iterations from an outer loop to an inner loop and then unroll the iterations (it will also do normal unrolling for any statements around the inner loop). The following is the result of applying this to `sgmev_n` (see Figure 7a top for the starting object code) with $c = 2$:

```
def sgemv_n(...):
    for io in seq(0, M / 2):
        for j in seq(0, N): # <--- main inner loop
            y[0+2*io] += x[j] * A[0+2*io, j]
            y[1+2*io] += x[j] * A[1+2*io, j]
    for ii in seq(0, M % 2):
        for j in seq(0, N): # <--- tail inner loop
            y[ii+M/2*2] += x[j] * A[ii+M/2*2, j]
```

In the first inner loop, `x[j]` is loaded from memory twice. To optimize this, we can load `x[j]` once and store it in a register for both updates, which is equivalent to applying CSE. Our implementation treats the resulting inner loop as a level-1 problem and calls `optimize_level_1` from Section 6.2.1.

***Triangular Matrix.*** In triangular matrices, the inner loop bounds depend on the outer loop variable, preventing direct loop reordering for data reuse. To address this, the inner loop bound is rounded up or down to a multiple of the blocking factor, removing the dependence on the outer loop variable. This allows the loops to be reordered using unroll-and-jam to enable data reuse optimizations. The exact rounding strategy depends on whether the diagonal is included and if the target machine supports predicated vector memory operations.

The combined scheduling code for general and triangular matrices, along with their performance graphs, can be found in Appendix D.2 [32] and in more detail in [19]. We optimized 50 kernel variants, supporting most BLAS level-2 operations (excluding banded operations and packed-triangular formats) across all configurations. This includes operations (gemv, ger, symv, syr, syr2, trmv, trsv), precisions (float (`s`), double (`d`)), operational parameters (transpose (`t`), non-transpose (`n`), lower (`l`), upper (`u`) triangular, unit (`u`), non-unit (`n`)), and target architectures (AVX2, AVX512). Our implementation achieved competitive performance with OpenBLAS, MKL, and BLIS on both AVX2 and AVX512 platforms.

***Skinny Matrix.*** A more efficient strategy for short vectors and skinny matrices is to load the entire vector into registers before the quadratic math loop. Since the vector is stored in registers, the vector length must be statically determined as the closest multiple of the register width, and specialized cases must be generated for each multiple.

Figure 7 shows the implementation of this shared schedule (7b) and its application on two programs (7a top) and (7c top). We developed this shared schedule for CPUs with vector extensions that support predicated vector loads/stores (e.g. x86 AVX2 and AVX512). The schedule follows four steps: (1) Inspect the program to obtain cursors to the inner loop (line 3) and the reused vector (line 4). (2) Stage the reused vector into registers by rounding up the loop iteration bound to a multiple of the vector width (line 5) and staging the vector around the doubly nested loops (line 6). The bottom of Figures 7a and 7c show the resulting object codes after this step. (3) Optimize the generated loops (load, inner math loop, and store) by applying vectorization (line 14) and interleaving the inner loop to increase ILP. (4) Specialize the program to ensure static information about the number of registers and their accesses is known for efficient code generation.

We compared the performance of the generated programs against Intel's MKL, OpenBLAS, and BLIS on problems with $N = 40$ and $M$ having powers of 2 and 3. Figure 8 shows the results, where each heatmap cell represents the geometric mean over problems in the respective x-axis bucket. The evaluated programs also handle scaling, which was omitted for brevity in Figure 7.

#### 6.2.3 Matrix-Matrix multiply.

The leading gemm literature, such as GotoBLAS [23] and BLIS [69], suggests building a general high-performance GEMM out of highly performant register-level tiles (micro-kernels). In addition to the main micro-kernel that fully utilizes registers and L1 cache size, we

```
def sgemv_n(M:size,N:size,A:[f32][M,N],
            x: [f32][N], y: [f32][M]):
  assert N <= 88
  for i in seq(0, M):
    for j in seq(0, N):
      y[i] += x[j] * A[i, j]
```

```
def sgemv_n(...):
  assert N <= 88
  var0: f32[(7 + N) / 8, 8] @ VEC_AVX2
  for i0 in seq(0, (7 + N) / 8 * 8):
    if i0 < N:
      var0[i0 / 8, i0 % 8] = x[i0]
  for i in seq(0, M):
    for j in seq(0, (7+N)/8*8):
      if j < N:
        y[i] += var0[j/8, j%8] * A[i, j]
```

**(a)** The object code of single-precision non-transposed matrix-vector multiplication.

```
1  def opt_skinny(p,
2                 out_loop, vw, mem, ...):
3    in_loop = get_inner_loop(p, out_loop)
4    vec = get_reused_vector(p, in_loop)
5    p = round_loop(p, in_loop, vw, up=True)
6    p, cs = auto_stage_mem(p, out_loop,
7                 vec.name(), rc=True)
8    p = set_memory(p, cs.alloc, mem)
9    p = divide_dim(p, cs.alloc, 0, vw)
10   loops = [cs.load,
11            in_loop,
12            cs.store]
13   loops = filter_c(~is_invalid)(p, loops)
14   p = apply(vectorize)(p, loops, vw, ...)
15   p = interleave_loop(p, in_loop, ...)
16   p = specialize(p, p.body(), ...)
17   p = unroll_loops(simplify(p))
18   return cleanup(p)
```

**(b)** The shared schedule.

```
def dgemv_t(M:size,N:size,A:[f64][M,N],
            x: [f64][M], y: [f64][N]):
  assert N <= 44
  for i in seq(0, M):
    for j in seq(0, N):
      y[j] += x[i] * A[i, j]
```

```
def dgemv_t(...):
  assert N <= 44
  var1: f64[(3 + N) / 4, 4] @ VEC_AVX2
  for i0 in seq(0, (3 + N) / 4 * 4):
    if i0 < N:
      var1[i0 / 4, i0 % 4] = y[i0]
  for i in seq(0, M):
    for j in seq(0, (3 + N) / 4 * 4):
      if j < N:
        var1[j/4, j%4] += x[i] * A[i, j]
  for i0 in seq(0, (3 + N) / 4 * 4):
    if i0 < N:
      y[i0] = var1[i0 / 4, i0 % 4]
```

**(c)** The object code of double-precision transposed matrix-vector multiplication.

**Figure 7.** 7b encapsulates BLAS level 2, skinny matrix-specific optimization in a single library function, used to optimize both transpose (7c) and non-transpose (7a) gemv and ger, amortizing the cost of writing schedules across kernel variants.

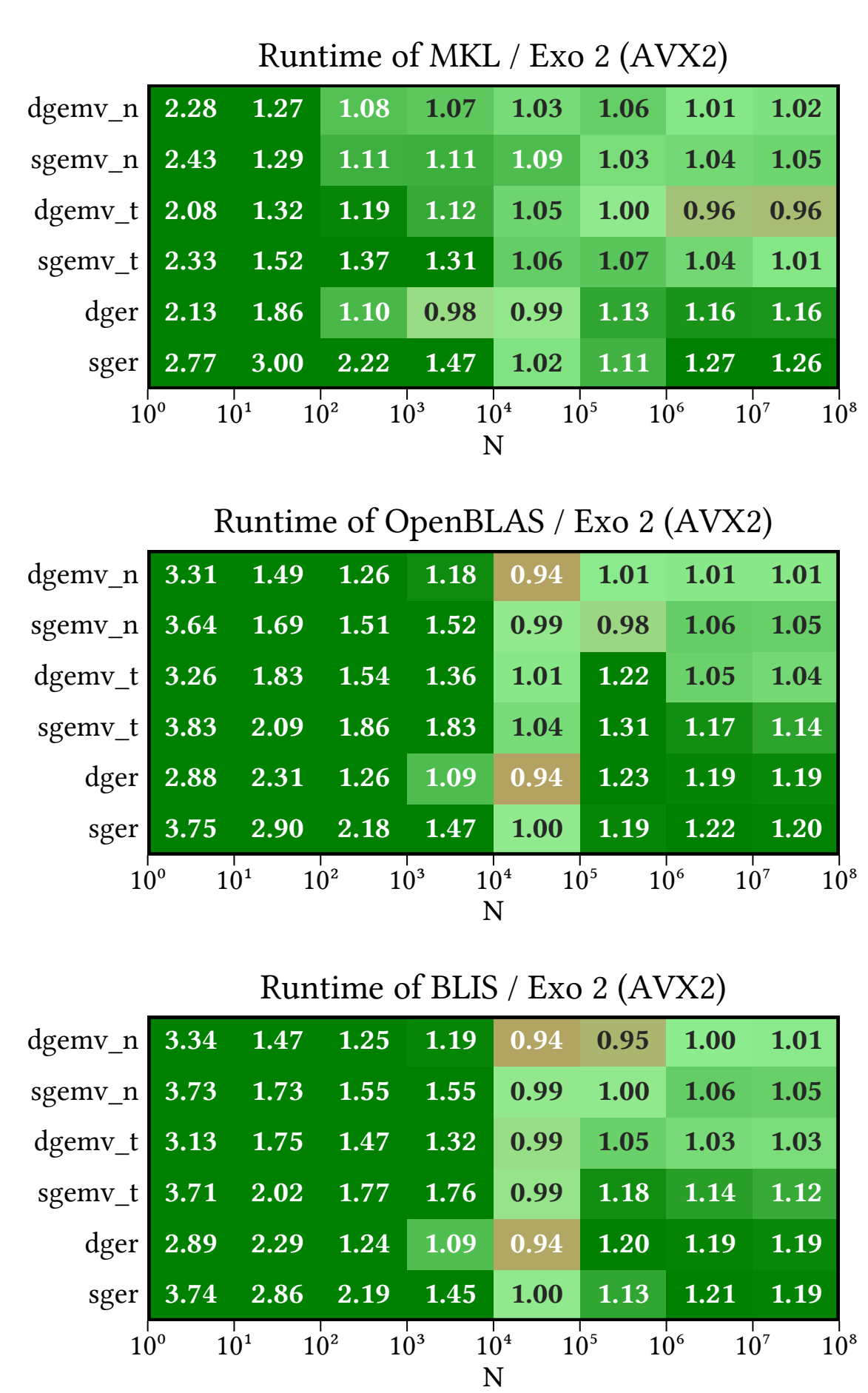

**Figure 8.** Performance evaluation of Exo 2 against Intel MKL, OpenBLAS, and BLIS on gemv and ger variants. `_n` suffix means non-transpose, and `_t` means transpose.

| | BLAS-lib | std-lib | ins-lib | Level 1 | Level 2 | sgemm |
|---|---|---|---|---|---|---|
| Obj. | 0 | 0 | 0 | 90 | 139 | 11 |
| Schd. | 241 | 1089 | 449 | 18 | 34 | 97 |
| Gen. C | N/A | N/A | N/A | 3196 | 11040 | 521 |

**(a)** Breakdown of the lines of code distribution in our library and kernels. std-lib refers to higher-order and target-specific functions like `vectorize`, BLAS-lib represents BLAS-specific optimizations such as `opt_skinny` and `optimize_level1`, and ins-lib refers to the inspection library functions.

| kernel | rewrites | kernel | rewrites | kernel | rewrites |
|---|---|---|---|---|---|
| asum | 1760 | axpy | 1826 | dot | 1208 |
| rot | 1716 | rotm | 3372 | scal | 1308 |
| swap | 1484 | gemv | 3996 | ger | 2848 |
| symv | 3310 | syr2 | 7832 | syr | 4046 |
| trmv | 4828 | trsv | 5370 | sgemm | 756 |

**(b)** Number of primitive rewrites required for optimizing each kernel. This includes the generation of all configurations, such as single and double precision, transpose and non-transpose operations, stride-1 and stride-any cases, and other variations (e.g., upper or lower triangular, and unit or non-unit options).

**Figure 9.** (a) Lines of code and (b) the number of primitive rewrites for library functions and kernels from Section 6.2. sgemm refers to the AVX512 sgemm kernel.

also need to generate smaller micro-kernels of size $m \times 16n$, where either $m = 6$ and $n \leq 4$, or $m \leq 6$ and $n = 4$. Generating these micro-kernels by hand is laborious and typically written in assembly in the aforementioned literature.

We implemented a single scheduling function, `gen_ukernel`, for all the micro-kernel generation, parameterized by precision, vector width, vector predicates, and the micro-kernel sizes $N$ and $M$. The `gen_ukernel` function stages memory for registers, applies `vectorize` (see Section 6.1.1), and inserts calls to AVX512 intrinsics. This function is used to generate all the micro-kernels in the main, right, and bottom panels (see Appendix C [32]).

Figure 9 shows the lines of code in the scheduling library and the number of primitive rewrites performed to optimize each kernel. Exo 2 reduces scheduling complexity by allowing users to implement new scheduling operators and reuse them for the optimization of multiple kernels rather than hand-writing primitives for each kernel one by one. We show the number of primitive rewrites required for scheduling kernels as it reflects what users would directly write in Exo. This may slightly overstate the difference compared to an equivalent best-effort Exo schedule, but accurately reflects the order of magnitude improvement we observe with Exo 2. Moreover, we compare the performance of the Exo 2 generated GEMM against the Exo generated GEMM. As shown in Figure 6, our implementation achieved similar or faster runtime compared to Exo's implementation (which is comparable to MKL's) while requiring fewer lines of code.

### 6.3 Reproducing Existing Scheduling Languages

Existing USLs hardcode actions and referencing schemes, limiting users to specific design choices. For instance, ELEVATE limits users to a linear time model and actions controlled by traversal functions. Halide limits users to a well-chosen set of scheduling primitives and a fixed-time, nominal referencing scheme. In contrast, Exo 2 allows users to choose referencing schemes and the level of automation in scheduling operators. By reproducing ELEVATE-style traversal functions, Halide's scheduling operations, and their respective referencing schemes within the user-level library code, we demonstrate that Exo 2 can recreate even complex actions such as `compute_at` and `store_at`, and show how Exo 2's branching time model encompasses the linear and fixed time reference models employed by other USLs. Our approach gives users not only the choice of operators and references but also enables interoperability between different user-defined operators and referencing schemes within the same program as needed.

#### 6.3.1 Reproducing ELEVATE.

***Traversal functions.*** Traversal functions specify a traversal order starting from a cursor and have type $\text{Top} = \text{Cursor} \rightarrow \text{Stream}[\text{Cursor}]$. In rewrite-based USLs, users must explicitly control the order in which scheduling actions are applied to the object code. This is because the sequence of actions does not commute in general, even when the same scheduling primitive is repeatedly applied, similar to the phase-ordering problem in general compilers. For example, suppose there is a three-nested loop of `i, j, k`. Applying `lift_scope` on `j` and then `k` will yield a loop nest of `j, k, i`, while applying it on `k` and then `j` will yield a loop nest of `i, j, k`. Therefore, it is necessary to define rules for different traversal strategies.

Strategy languages [43, 68] and USLs like ELEVATE separate traversal strategies from rewrite rules. By leveraging navigation and introspection, we can support traversal strategies over cursors, such as the postorder traversal (`lrn`).

```
def lrn(c):
  for c in c.body():
    if isinstance(c,
        (ForCursor,IfCursor)):
      yield from lrn(c)
    yield c
```

***Linear time.*** Moreover, we can reproduce a functional-style linear-time referencing model using higher-order functions. Cursors were implicitly forwarded and returned in all the higher-order scheduling functions we discussed earlier (see Section 3.4). However, we can also define higher-order scheduling functions that return other cursors by using the `nav` function, which navigates the frame of reference:

```
def nav(move):
  return lambda p,c: p, move(p.forward(c))
def savec(op):
  return lambda p,c: op(p,c)[0], c
```

Let `move` : Cursor → Cursor be a function that transforms a cursor. `savec` allows us to navigate the reference frame arbitrarily within the argument `op`, but restore it afterwards. We can compose these combinators into a useful pattern where the cursor is navigated to take some action and then restored afterward. This pattern allows us to avoid ambiguity in our references by staging all operations relative to a stable point within the AST.

```
reframe = lambda move, op: savec(seq(nav(move), op))
```

Because `nav` forwards the cursor before performing any spatial navigation, these combinators have now recreated linear-time reference frames (Section 5.1) used by strategy languages and ELEVATE. Furthermore, as shown below, Exo 2 concisely recreates many Exo scheduling primitives that combined actions with relative references, in a single line.

```
reorder_before     = reframe(lambda c:c.expand(1, 0),
                             lift(reorder_stmts))
remove_parent_loop = reframe(lambda c:c.parent(),
                             lift(remove_loop))
fission_after      = reframe(lambda c:c.after(),
                             lift(fission))
```

#### 6.3.2 Reproducing Halide.

***Blur in Halide and Exo.*** As shown in Figure 11, the 3x3 blur algorithm and schedule from the Halide paper [56] uses built-in actions (`compute_at`, `tile`, `vectorize`, `parallel`) and nominal referencing. When `compute_at` is called without a corresponding `store_at`, Halide automatically stores the buffer at the same loop level as the computation. The equivalent Exo starting object code (Figure 10 upper left, after tiling to `y` and `x` has already been applied) is longer than the Halide version because Exo's lower-level IR explicitly expresses loops, allocations, and read/write/reduce statements, while Halide is restricted to pure functions of integer coordinates. For simplicity, we restrict input images to whole multiples of the tile size.

```
for y in seq(0, H+2):
 for x in seq(0, W+2):
  blur_x[y,x] = …
for y in seq(0, H/32):
 for x in seq(0, W/256):
  for yi in seq(0, 32):
   for xi in seq(0, 256):
    # Y = 32*y+yi, X = 256*x+xi
    blur_y[Y,X] = avg(blur_x[Y:Y+3,X])
```

Fuse over y loops →

```
for y in seq(0, H/32):
 for yi in seq(0, 34):
  for x in seq(0, W+2):
   blur_x[…] = …
 for x in seq(0, W/256):
  for yi in seq(0, 32):
   for xi in seq(0, 256):
    # Y = 32*y+yi, X = 256*x+xi
    blur_y[Y,X] = avg(blur_x[Y:Y+3,X])
```

Fuse over x loops →

```
for y in seq(0, H/32):
 for x in seq(0, W/256):
  for xi in seq(0, 258):
   for yi in seq(0, 34):
    blur_x[…] = …
  for yi in seq(0, 32):
   for xi in seq(0, 256):
    # Y = 32*y+yi, X = 256*x+xi
    blur_y[Y,X] = avg(blur_x[Y:Y+3,X])
```

**Inspection:** The outermost unfused loop of the producer (blur_x) is not the x loop

reorder_loops(p, loops) →

```
for y in seq(0, H/32):
 for x in seq(0, W+2):
  for yi in seq(0, 34):
   blur_x[…] = …
 for x in seq(0, W/256):
  for yi in seq(0, 32):
   for xi in seq(0, 256):
    blur_y[…] = …
```

**Inspection:** Within loop, we access the producer (blur_x) along the x dimension at (256*y, 256*y+258).

divide_with_recompute(p, loop, W+2, 256, ["x", "xi"]) →

```
for y in seq(0, H/32):
 for x in seq(0, W/256):
  for xi in seq(0, 258):
   for yi in seq(0, 34):
    blur_x[…] = …
 for x in seq(0, W/256):
  for yi in seq(0, 32):
   for xi in seq(0, 256):
    blur_y[…] = …
```

fuse(p, loop1, loop2) →

**Figure 10.** Exo 2 implementation of Halide's `blur_x.compute_at(blur_y, x)` operation. The starting code is a tiled version of the blur algorithm. We first fuse the `blur_x` and `blur_y` computations at the `y` loop level, and then fuse at the `x` loop level. Focusing on the second fusion, the bounds inference inspection described in Section 4 determines the producer (`blur_x`) accesses in the `x` loop. Then, a sequence of primitive actions is applied to the object code to yield code fused at the `x` loop level.

```
// ALGORITHM
blur_x(x,y)=(inp(x,y)+inp(x+1,y)+inp(x+2,y))/3;
blur_y(x,y)=(blur_x(x,y)+blur_x(x,y+1)+blur_x(x,y+2))/3;
// SCHEDULE
blur_y.tile(x, y, xi, yi, 256, 32)
      .vectorize(xi, 16)
      .parallel(y);
blur_x.compute_at(blur_y, x)
      .vectorize(x, 16);
```

**Figure 11.** Halide's blur algorithm and schedule.

```
p = H_tile(p,"blur_y","y","x","yi","xi",32,256)
p = H_compute_store_at(p,"blur_x","blur_y","x")
p = H_parallel(p, "y")
p = H_vectorize(p, "blur_x", "xi", 16)
p = H_vectorize(p, "blur_y", "xi", 16)
p = H_store_in(p, "blur_x", DRAM_STACK)
```

**Figure 12.** Exo 2's blur schedule.

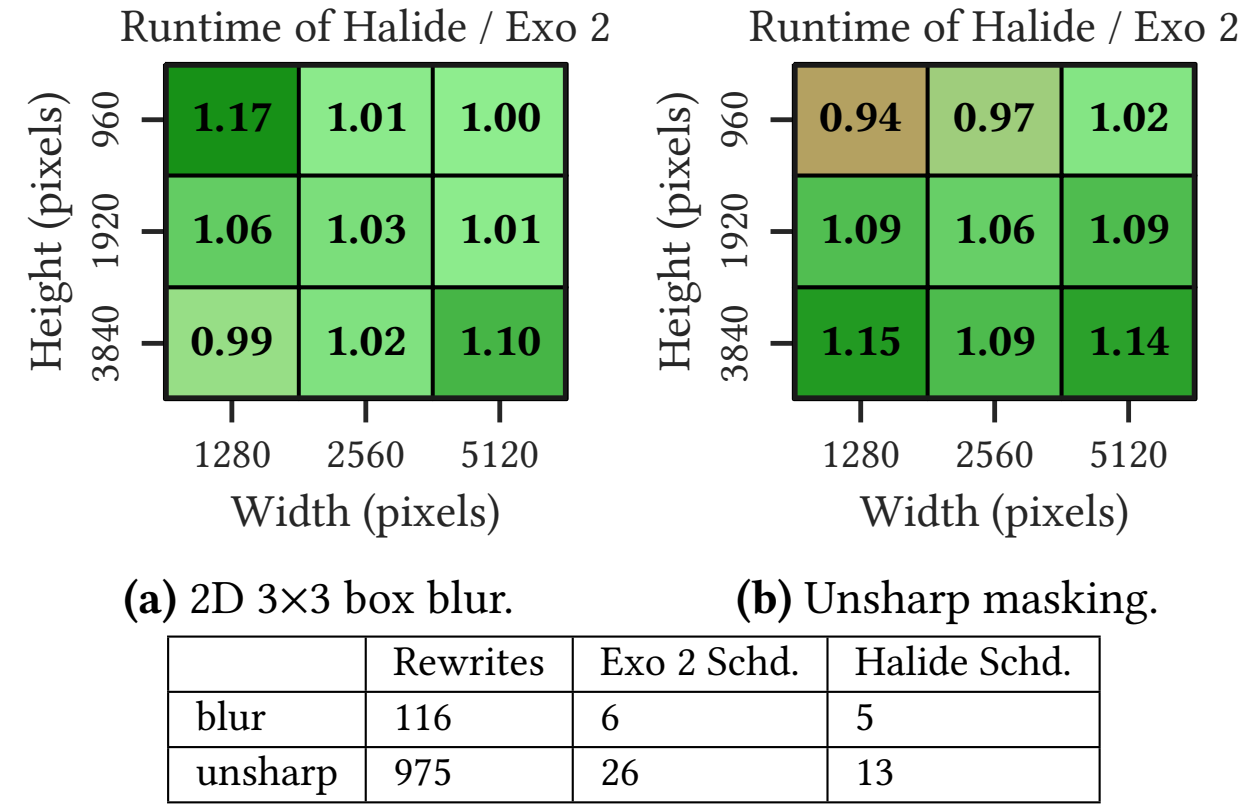

**(a)** 2D 3×3 box blur. **(b)** Unsharp masking.

| | Rewrites | Exo 2 Schd. | Halide Schd. |
|---|---|---|---|
| blur | 116 | 6 | 5 |
| unsharp | 975 | 26 | 13 |

**(c)** Lines of code for the Exo 2 schedule, Halide schedule, and the number of rewrites performed during the Exo 2 scheduling process.

**Figure 13.** Exo 2 schedules show competitive performance to corresponding expert-written Halide schedules on a variety of image sizes.

***Exo 2 Blur Schedule.*** Halide's `compute_at` and `store_at` scheduling operations use automated bounds inference to determine loop bounds and buffer sizes. We leverage the previously discussed bounds inference inspection (Section 4) to reproduce these Halide operations. Figure 10 summarizes our `compute_at` implementation, and similar ideas apply to `store_at`. To bridge the gap between Halide's and Exo 2's referencing schemes, we defined `H_`-prefixed functions that expect nominal references and internally convert to Exo 2's cursor references. Halide's nominal referencing scheme uses action-specialized built-in navigation. For example, `blur_x.compute_at(blur_y, x)` refers to the `x` loop enclosing the `blur_y` computation. Since cursors generalize nominal referencing, this simply involves translating the built-in navigation performed by Halide's scheduling operations. As shown in Figure 12, using the Halide-like scheduling operations with nominal referencing, Exo 2 can reproduce Halide-style schedules. `H_compute_store_at` is called to implement Halide's aforementioned compile-time decision of calling both `compute_at` and `store_at`.

With this interface, we optimized 3×3 box blur and unsharp masking and compared the results with expert-written Halide schedules [14]. Figure 13 shows that Exo 2's Halide-like schedules offer similar performance to Halide. More details about Exo 2's Halide library can be found in [53].

## 7 Related Work

***Homogeneous and heterogeneous meta-programming.*** Scheduling is fundamentally a meta-programming process where the schedule is a meta-program that derives optimized implementations from a base object program. Tim Sheard introduced a distinction between homogeneous and heterogeneous meta-programming systems: homogeneous systems use the same language for both meta-programs and object programs, whereas heterogeneous systems use different languages for the two [60]. Staged meta-programming has been extensively explored in homogeneous systems for code generation and building DSLs since the 1990s, pioneered by languages such as MetaML, MetaHaskell, MetaOCaml, Lisp, and Racket [20, 38, 45, 46, 61, 62, 65, 66], and remains an active research area in modern systems [6, 8, 51, 74]. The Terra-Lua system [15, 16] introduced a heterogeneous meta-programming approach, using Lua [30], a dynamic, high-level language, to meta-program Terra, an embedded, lower-level object language. USLs are heterogeneous systems since the algorithm and scheduling languages have separate syntax and semantics, but are fundamentally different from Terra because they are *transformative.*

***Generative and transformative meta-programming.*** We propose a new taxonomy within meta-programming languages, distinguishing between *generative* and *transformative* meta-programming. Generative meta-programs create new object code by constructing and interpolating code fragments, while transformative meta-programs deconstruct and transform a code object into another code object. Lisp-inspired meta-programming systems and Terra are predominantly generative, creating new code at compile-time using constructs like `quote` and `unquote`. On the other hand, jQuery [36], XSLT [12], ELEVATE [26], and its predecessor strategy languages [29, 43, 68] focus on transformative meta-programming. jQuery enables users to navigate and manipulate the HTML DOM using methods like `.parent()` and `.prev()`, while XSLT allows users to transform XML documents via XPath pattern matching. ELEVATE and strategy languages use pattern matching to obtain references to the object code. Racket, with `syntax-case` and `syntax-parse`, supports both generative and transformative meta-programming by allowing code transformations while preserving lexical scoping and hygiene. Inspired by transformative approaches, Exo 2 supports jQuery-like navigation and a `.find()` operation (Section 5). USLs are a subset of transformative systems which preserve functional equivalence.

***User-schedulable languages.*** Halide [54, 55] popularized the idea of USLs, and subsequent Halide-inspired languages explored different target applications and control-abstraction boundaries [7, 10, 25, 28, 31, 39, 64, 70, 71, 75]. For instance, TVM offers accelerator abstraction through `tensorize`, TACO separates tensor definitions from sparse formats, and Taichi abstracts memory layout and management with `fields`. We highlighted the limitations of existing USLs, including their lack of inspection capabilities (Section 4), inflexible "nominal" referencing schemes, and lack of support for multiple, time-stable references (Section 5).

The Transform dialect [44] is a language used to transform one MLIR program to another through a transformative meta-programming. Although it does not guarantee the safety of its transformations, its interface is similar to other USLs and provides a reference mechanism called *handles.* In our terminology, handles are *one-time, nominal* references. Handles are labels placed on the object IR, rather than a mechanism for navigating the code, since handles are invalidated after any transformation and have to be reassigned a new name. This is consistent with the SSA nature of MLIR.

Exo [31] introduced the *exocompilation* paradigm by externalizing hardware-specific definitions to user code and offering low-level primitives for precise performance control. Exo 2 is a new iteration of Exo, designed to overcome limitations we encountered while developing a variety of kernels in Exo, particularly the challenge of sharing common scheduling operations across different applications. Unlike Exo and other existing USLs that allow users to write new *schedules*, Exo 2 also enables users to write new *scheduling operators*, enabling the creation of reusable scheduling libraries. Key technical differences lie in the cursor, inspection, and reference mechanisms. Exo's only reference mechanism is structural pattern-matching (e.g., `"for i in _:_"`) which (like MLIR handles) are *one-time, nominal* references, and lack support for cursor forwarding, relative navigation, and object code inspection.

The fundamental problem with Exo and exiting USLs is that it is difficult or impossible to parameterize scheduling library functions based on the references on which they operate, making it infeasible to build scheduling libraries. Specifically, encapsulating user-defined functions into libraries of *new scheduling operators*, as showed in Section 6, is not possible with existing USLs. Therefore, their schedules are often formed out of repeated, explicit instantiations of scheduling patterns, rather than simple invocations to encapsulated generalizations of those patterns. Such code is more difficult to maintain and modify for new purposes. Inspired by Guy Steele's famous talk "Growing a Language," in which he argues that a language should not be static but should provide mechanisms for users to introduce new abstractions and functionality, we set out to design a scheduling language that can scale to large projects through libraries of user-defined scheduling operators [63].

Auto-schedulers for Halide [1, 4, 49] and TVM [11, 59, 76] automate the scheduling process for USLs, while equality saturation [67, 72] automatically explores program optimizations. Guided optimization [33] and guided equality saturation [41] have been proposed to improve these automatic techniques by incorporating user guidance. Similarly, Exo 2 believes user input is necessary for scheduling, but takes a bottom-up approach by building automation from fine-grained primitives and allowing users to intervene as needed, in contrast to the top-down approach of guided techniques.

***Theorem provers.*** Our bottom-up approach to building safe scheduling abstractions is inspired by theorem provers. In the 1970s, to address the issue of LCF [35] sometimes combining incorrect proofs and asserting their validity [48, 52], Robin Milner developed ML to only allow users to construct valid proofs within the language using a set of trusted axioms [47]. This method, known as "the LCF approach," has led to modern theorem provers like Coq [22]. Inspired by this, Exo 2 enables the creation of complex scheduling programs by composing trusted, equivalence-preserving primitives.

***HPC libraries.*** BLAS pioneered the standardization of HPC kernels [17, 18, 27, 42]. Many BLAS libraries have been developed by hardware vendors, optimized for their particular hardware, such as Intel's oneMKL [34], Apple's Accelerate Framework [5], and Nvidia's cuBLAS [50]. Open-source implementations like ATLAS [13], GotoBLAS [23, 24] (now OpenBLAS [73]), and BLIS [69] have introduced various techniques for high-performance, portable BLAS implementations. Previous attempts to generate GEMM microkernels using Exo led to schedules with limited parametrization [9] due to the aforementioned Exo's limitations.

Unlike BLIS, which focuses on reusing *optimized object code* (microkernel) across different kernels, Exo 2 enables a library design that reuses *optimization strategies* as parameterized meta-programming functions that transform the object code. This allows for another layer of code reuse across different kernels. For instance, BLIS's microkernel is hand-written in assembly for each size, hardware target, and precision. In contrast, as shown in Section 6.2, our library written in Exo 2 can parameterize the microkernel generation and generate these variants from a *single* library function. Moreover, the BLAS level 1 library function could be reused for optimizing the inner loop of level 2 kernels, as the level 2 inner loop shares the same optimization strategy as level 1 (Section 6.2.2). Such reuse of scheduling strategies is impossible without meta-programming capabilities.

## 8 Discussion

While Exo 2 is designed for growth, it does not address the problem of expressibility in USLs; the finest level of control provided is still dictated by the set of scheduling primitives. If no composition of primitives achieves a desired transformation, modifying the compiler remains necessary. However, ever finding a complete set of primitives seems unlikely, as scheduling operations represent axioms of program equivalence, an inherently undecidable problem. Nonetheless, our choice of Exo 2 primitives was expressive enough to build libraries that achieved performance on par with state-of-the-art HPC libraries on multiple platforms.

Composing many fine-grained primitives that guarantee functional equivalence through static analysis can increase compilation time compared to higher-level, built-in operations. This was especially evident in more complex schedules such as GEMM and unsharp masking, which took 30 seconds and 2 minutes to schedule, respectively. Additional sources of overhead include our reliance on SMT solvers for checking functional equivalence and the fact that Exo 2 is implemented in Python. We plan to improve Exo 2's scheduling time in future work through both caching and more efficient analysis algorithms.

The primary contribution of this paper is making it possible to "build up" libraries of scheduling abstractions – not just individual schedules – from fine-grained primitive operations. We believe this is essential to the design of better USLs in general. Our secondary contributions include conceptualizing scheduling languages through the action, inspection, and reference framework, and introducing the Cursor mechanism. These contributions serve as a means to achieve our primary goal of building *new scheduling operators* from fine-grained primitives.

We hope that empowering users to create automation in libraries external to the compiler inspires a new workflow for writing high-performance code. Instead of performance engineers and HPC library authors tediously optimizing kernels one by one while fixing correctness bugs, we argue that a more efficient library design should separate the process into three responsibilities: *USL designers* provide primitives and ensure their safety and correctness; *scheduling library authors* write target- and application-specific optimization strategies; and *HPC library authors* effectively optimize kernels using these library functions.

## Acknowledgments

We would like to express our deepest gratitude to all the early users of Exo who provided us with valuable feedback that shaped this work. We are particularly thankful to David Akeley, Rahul Bakshi, Manya Bansal, Julian Bellavita, William Brandon, Adrian Castello, Abhijit Davare, Sankalp Dayal, Zach DeVito, Grace Dinh, Asaf Hargil, Pritpal Kanhaiya, Amanda Liu, Aaftab Munshi, Adam Nemet, Rachit Nigam, Albert Ou, AJ Root, Ali Sazegari, and Philippe Tillet for their insightful discussions and early feedback. We also acknowledge the early adopters of Exo at the Vector and Numerics team at Apple and Amazon Devices. This research was funded in part by the U.S. Government under the DARPA RTML program (contract FA8650-20-2-7006) and NSF awards CCF-2328543 and TI-2303735. The first author was additionally supported by Masason, Funai, and Quad Fellowships.

# Artifact Appendix

## Abstract

The scripts and step-by-step guide to reproduce the figures in this paper are available at https://github.com/exo-lang/exo2-artifact. The Exo 2 compiler implementation is integrated into Exo at https://github.com/exo-lang/exo, and the Exo 2 BLAS library implementation is available at https://github.com/exo-lang/ExoBLAS.

## Artifact check-list (meta-information)

- **Binary:** Available as a PyPI package `exo-lang`
- **Run-time environment:** Tested on Arch Linux and Ubuntu 22.04.5 LTS
- **Hardware:** Intel machines with AVX512 and AVX2 instructions. An AWS FPGA instance is needed to run Firesim, which provides a cycle-accurate simulator for Gemmini.
- **Metrics:** Geomean speedup over existing libraries
- **How much disk space required (approximately)?:** Approximately 50GB of disk space is required. If simulating Gemmini on Firesim, 200GB of disk space is needed.
- **How much time is needed to prepare workflow (approximately)?:** 1 hour
- **How much time is needed to complete experiments (approximately)?:** 5 hours per existing library × hardware target combination
- **Publicly available?:** Yes
- **Code licenses (if publicly available)?:** MIT
- **Archived (provide DOI)?:** 10.5281/zenodo.13997025

## How to access.

- Artifact evaluation repository with step-by-step instructions: https://github.com/exo-lang/exo2-artifact
- Compiler: https://github.com/exo-lang/exo
- BLAS library: https://github.com/exo-lang/ExoBLAS

## Software dependencies.

- Python>=3.9
- To reproduce Halide results: Halide 16.0.0
- To reproduce BLAS results: cmake>=3.23, Ninja, Google Benchmark, and one or more existing BLAS libraries (MKL, OpenBLAS, or BLIS). Also, install the Python packages listed in `requirements.txt` from the ExoBLAS repository.
- To reproduce Gemmini results: Firesim and Chipyard

## Installation

You can either build from the source or install from PyPI:

```
$ pip install exo-lang
```

## Experiment workflow

We will focus on explaining the BLAS results (Figures 8, 9, 14, 15, 16, 17, 18, and 19) to demonstrate the workflow, but the full guide can be found in the repository. After installing the requirements, navigate to the ExoBLAS repository and run the following commands to build the library for the AVX512 target:

```
$ cmake --preset avx512
$ cmake --build build/avx512/
```

To run the benchmarks:

```
$ ctest --test-dir ./build/avx512 -R bench
```

The performance graphs in the paper can be reproduced using the following commands:

```
$ ./analytics_tools/graphing/organize.sh
↪  benchmark_results
$ python3 ./analytics_tools/graphing/graph.py all
↪  AVX512 benchmark_results/level1
```

Change `level1` to `level2` above to generate level 2 graphs.

## Evaluation and expected results

After running the above scripts, you should see the generated graphs in the `analytics_tools/graphing/graphs` directory as PDF files. Each graph includes all the heatmap graphs for the specified hardware target, comparing the performance of the Exo 2 library against the respective reference BLAS implementation.

## Experiment customization

Code generation for different hardware targets and linkage to other BLAS libraries can be controlled by modifying the `cmake -preset` command. To generate code for the AVX2 instruction set, change the CMake preset accordingly.

```
$ cmake --preset avx2
```

If unspecified, CMake will attempt to find an existing BLAS implementation to link against. If you wish to control which existing library to compare the performance against, you can use the `-DBLA_VENDOR` option as follows:

```
# Use OpenBLAS as a reference
$ cmake --preset avx512 -DBLA_VENDOR=OpenBLAS
# Use MKL as a reference
$ cmake --preset avx512 -DBLA_VENDOR=Intel10_64lp_seq
# Use BLIS as a reference
$ cmake --preset avx512 -DBLA_VENDOR=FLAME
```

# A List of Scheduling Primitives

## A.1 Loop Transformations

| Scheduling operation | Code transformation | Safety conditions |
| --- | --- | --- |
| reorder_loops(p, loops) | for i:<br>  for j:<br>    s<br>⇝<br>for j:<br>  for i:<br>    s | The j loop's bounds cannot depend on i. The loop body s commutes for different $(i, j)$ and $(i', j')$ pairs. |
| divide_loop(<br>  p,<br>  loop,<br>  c,<br>  [io, ii],<br>  tail_strategy<br>) | # tail_strategy=perfect<br>for i < I:<br>  s<br>⇝<br>for io < I/c:<br>  for ii < c:<br>    s[i ↦ c*io+ii]<br># tail_strategy=guard<br>for i < I:<br>  s<br>⇝<br>for io < I/c:<br>  for ii < c:<br>    if c*io+ii < I:<br>      s[i ↦ c*io+ii]<br># tail_strategy=cut<br>for i < I:<br>  s<br>⇝<br>for io < I/c:<br>  for ii < c:<br>    s[i ↦ c*io+ii]<br>for ii < I%c:<br>  s[i ↦ c*I/c+ii]<br># tail_strategy=cut_and_guard<br>for i < I:<br>  s<br>⇝<br>for io < I/c:<br>  for ii<c:<br>    s[i ↦ c*io+ii]<br>if I%c > 0:<br>  for ii < I%c:<br>    s[i ↦ c*I/c+ii] | For perfect tail_strategy, the loop bound is perfectly divisible by c. |
| divide_with_recompute(<br>  p, loop, N, c, [io, ii]<br>) | for i < I:<br>  s<br>⇝<br>for io < N:<br>  for ii < c+I-N*c:<br>    s[i ↦ c*io+ii] | s is idempotent and N*c ≤ I. |
| mult_loops(p, loops, k) | for i < I:<br>  for j < c:<br>    s<br>⇝<br>for k < I*c:<br>  s[i ↦ k/c, j ↦ k%c] | The j loop is the only statement in the i loop's body. Also, c is constant. |
| cut_loop(p,loop,e) | for i in l, h:<br>  s<br>⇝<br>for i in l, e:<br>  s<br>for i in e, h:<br>  s | The cutoff e lies between the loop bounds l and h |
| join_loops(p, loop, loop2) | for i in l1, h1:<br>  s<br>for i2 in l2, h2:<br>  s<br>⇝<br>for i in l1, h2:<br>  s | The loops are adjacent, have identical bodies, and h1 = l2. |
| shift_loop(p, loop, e) | for i in l, h:<br>  s<br>⇝<br>for i in e, h+e-l:<br>  s | The new lower bound e $\geq 0$. |
| fission(p, gap) | for i:<br>  s1<br>  s2<br>⇝<br>for i:<br>  s1<br>for i:<br>  s2 | s2 cannot depend on allocations in s1, and effects of s1 and s2 commute |
| remove_loop(p, loop) | for i:<br>  s<br>⇝<br>s | s is idempotent and cannot depend on i. Loop executes at least once. |
| add_loop(p, s, i, hi, guard) | # guard = False<br>s<br>⇝<br>for i < hi:<br>  s<br># guard = True<br>s<br>⇝<br>for i < hi:<br>  if i == 0:<br>    s | s is idempotent and hi is positive. |
| unroll_loop(p, loop) | for i in lo, hi:<br>  s<br>⇝<br>s[i ↦ lo]<br>...<br>s[i ↦ hi - 1] | hi and lo are constants, hi - lo $> 0$ |

## A.2 Code Rearrangement

| Scheduling operation | Code transformation | Safety conditions |
|---|---|---|
| `reorder_stmts(p, s1, s2)` | s1<br>s2 ⇝ s2<br>s1 | s1 and s2 commute. |
| `commute_expr(p, e)` | x+y ⇝ y+x<br>x*y ⇝ y*x | N/A |

## A.3 Scope Transformations

| Scheduling operation | Code transformation | Safety conditions |
|---|---|---|
| `specialize(p, s, conds)` | s ⇝ if conds[0]:<br>  s<br>else:<br>  if conds[1]:<br>    s<br>  else:<br>    ... | conds only contains valid boolean expressions. |
| `fuse(p, scope, scope2)` | for i<I:<br>  s<br>for i<I:<br>  s2<br>⇝<br>for i<I:<br>  s<br>  s2 | Equivalent upper loop bound I in its context. Iterations of s commute with any iteration of s2. |
| | if e:<br>  s<br>if e:<br>  s2<br>⇝<br>if e:<br>  s<br>  s2 | Equivalent if conditional expression e in its context. |
| `lift_scope(p, scope)` | | For all cases, scope is the only statement in its parent's body. |
| | if e:<br>  if e2: # scope<br>    s<br>  else:<br>    s2<br>else:<br>  s3<br>⇝<br>if e2:<br>  if e:<br>    s<br>  else:<br>    s3<br>else:<br>  if e:<br>    s2<br>  else:<br>    s3 | N/A |
| | if e: # scope<br>  for i:<br>⇝<br>for i:<br>  if e: | if statement cannot have an else clause. |
| | for i:<br>  if e: # scope<br>    s<br>  else:<br>    s2<br>⇝<br>if e:<br>  for i:<br>    s<br>else:<br>  for i:<br>    s2 | e cannot depend on i. |

## A.4 Multiple procedures

| Scheduling operation | Code transformation | Safety conditions |
|---|---|---|
| `inline(p, foo)` | Inline a callsite of foo | N/A |
| `replace(p,s,instr)` | Replace s with a call to an instruction instr | s and the instr function are unifiable. |
| `call_eqv(p,foo,bar)` | Replace foo with a call to an equivalent procedure bar | The two procedures foo and bar are equivalent, e.g. scheduled from the same procedure |
| `extract_subproc(p,s,foo)` | Isolates s into a function named foo, and replaces s with a call to foo. | N/A |

## A.5 Buffer Transformations

| Scheduling operation | Code transformation | Safety conditions |
|---|---|---|
| lift_alloc(p, a) | for i:<br>  a: T[sz]<br>  s<br>⇝<br>a: T[sz]<br>for i:<br>  s | Dimensions sz don't depend on i. No loop-carry dependencies. |
| sink_alloc(p, a) | a: T<br>for i:<br>  s<br>⇝<br>for i:<br>  a: T<br>  s | a is only accessed in the i loop. No loop-carried dependencies. |
| delete_buffer(p, a) | a: T<br>s<br>⇝<br>s | a should be dead. |
| reuse_buffer(p, a, b) | a : T[sz]<br>...<br>b : T[sz]<br>s<br>⇝<br>a : T[sz]<br>...<br>s[b ↦ a] | a and b have the same type and size. a is dead after b's allocation. |
| resize_dim(<br>  p, a, dim, sz, off, fold<br>) | ## fold = False<br>a: T[_]<br>s<br>⇝<br>a: T[_]<br>s[a[e] ↦ a[e-off]] | l < h, a only accessed in dim-th dimension between l and h. |
| | ## fold = True<br>a: T[sz]<br>s<br>⇝<br>a: T[sz]<br>s[a[e] ↦ a[(e-off)%sz]] | l < h, a only accessed in dim-th dimension between l and h. Also, accesses to a never access more than sz earlier than the largest access so far. |
| expand_dim(p, a, sz, e) | a: T[_]<br>s<br>⇝<br>a: T[_]<br>s[a[_] ↦ a[e, _]] | sz is positive, e only uses existing variables, and e < sz in all contexts. |
| rearrange_dim(p, a, p_vec) | # p_vec = [2, 0, 1]<br>a: T[N, M, K] ⇝ a: T[K, N, M] | a cannot be windowed or passed in function calls. |
| divide_dim(p, a, dim, c) | # dim = 0, c = 4<br>a: R[12, _]<br>s<br>⇝<br>a: R[3, 4, _]<br>s[a[i, _] ↦ a[i/4, i%4, _]] | a's dim-th dimension size is constant and divisible by c |
| mult_dim(p, a, dim, dim2) | # dim = 0, dim2 = 2<br>a: R[n, _, 4]<br>s<br>⇝<br>a: R[4*n,_]<br>s[a[i, _, j] ↦ a[4*i+j, _]] | a cannot be windowed or passed in function call. One of the dimensions is constant size. |
| unroll_buffer(p, a, dim) | a: T[c]<br>⇝<br>a1: T<br>...<br>ac: T | For this dimension, a has constant size and index accesses. a cannot be windowed along this dimension. |
| bind_expr(p, e, a, cse) | s<br>⇝<br>a: T<br>a = e<br>s[e ↦ a] | N/A |
| stage_mem(p, s, a, w, tmp) | # w = a[0:n, j-1:j]<br>for i < n-1: # s<br>  a[i,j] = a[i+1,j-1]<br>⇝<br>tmp: T[n, 2]<br>for k0 < n:<br>  for k1 < 2:<br>    tmp[k0, k1] = a[k0,j-1+k1]<br>for i < n-1:<br>  tmp[i, 1] = tmp[i+1, 0]<br>for k0 < n:<br>  for k1 < 2:<br>    a[k0, j-1+k1] = tmp[k0, k1] | The code block s does not access buffer a outside of the given window. |

## A.6 Simplification

| Scheduling operation | Code transformation | Safety conditions |
|---|---|---|
| simplify(p) | does arithmetic simplifications and trivial branch elimination on the entire procedure | N/A |
| eliminate_dead_code(p, scope) | for i in l, h: ⇝ pass | The loop runs 0 times, e.g. l ≥ h. |
| | if e:<br>  s<br>else<br>  s2 ⇝ s or s2 | e is equivalent to True or False in its context. |
| rewrite_expr(p, e, e') | e ⇝ e' | e and e' are equivalent in the context in which e appears. |
| merge_writes(p, s1, s2) | x = e1<br>x = e2 ⇝ x = e2 | The two statements write to the same destination x.<br>e2 cannot read from x. |
| | x += e1<br>x = e2 ⇝ x = e2 | |
| | x = e1<br>x += e2 ⇝ x = e1 + e2 | |
| | x += e1<br>x += e2 ⇝ x += e1 + e2 | |
| inline_window(p, w) | w = a[_]<br>s ⇝ s[w ↦ a[_]] | N/A |
| inline_assign(p, x = e) | x = e<br>s ⇝ s[x ↦ e] | x cannot be written to in the block s after s. |

## A.7 Backend-Checked Annotations

All the scheduling primitives that we have discussed so far are safety-checked within the their rewrite process. By contrast, consistency of precision types, memory annotations, and window annotations are performed as back-end checks after all scheduling is complete; immediately prior to code generation.

| Scheduling operation | Code transformation | Compilation safety check |
|---|---|---|
| set_memory(p, a, MEM') | a @ MEM ⇝ a @ MEM' | Ensures that the accesses to the buffer obeys the custom memory read/write/window definition, and the caller and callee have the same memory types |
| set_precision(p, a, T') | a: T ⇝ a: T' | Checks the caller, callee, and the both sides of binary operation have the same precision types |
| parallelize_loop(p, loop) | Annotate loop as parallel | Ensures that the loop iterations do not have RAW or WAW dependencies. |
| set_window(p, a) | a: T[_] ⇝ a: [T][_] | Checks the caller and callee have the same window shapes |

## A.8 Configuration State

| Scheduling operation | Code transformation | Safety conditions |
|---|---|---|
| bind_config(p, e, cfg, field) | s ⇝ cfg.field = e<br>s[e ↦ cfg.field] | cfg.field is not read by code which executes afterwards. |
| delete_config(p, cfg.field = _) | s<br>cfg.field = _ ⇝ s<br>s2              s2 | cfg.field is not read by code which executes afterwards. |
| write_config(p, gap, cfg, field, e) | s<br>s2 ⇝ s<br>cfg.field = e<br>s2 | cfg.field is not read by code which executes afterwards. |

## B Matrix-multiply on Gemmini

This section provides a step-by-step guide to optimizing matrix multiplication for the Gemmini hardware accelerator. We start with a simple GEMM object code and progressively transform it to leverage Gemmini's unique hardware features. The code snippets on the right illustrate both the object and the scheduling code at each stage of the optimization process. Exo object code is enclosed in boxes, while Exo 2 scheduling code has a vertical line on the left. Most of the scheduling functions used in this example are written in user-code, as part of Exo 2's Gemmini library.

***Initial Object Code.*** The starting point is a standard GEMM object code with additional post-processing steps specific to machine learning applications. These steps include scaling, clamping, and applying a ReLU activation function before storing the final result. The scaling and clamping operations ensure correct quantization for integer matrix multiplication, while the activation function is a common component in machine learning pipelines. Compared to Exo's PLDI'22 artifact, our initial algorithm is two lines shorter due to the removal of unnecessary staging of input matrices `A` and `B`.

```
def matmul_on_gemmini(N: size, M: size, scale: f32,
    act:bool, A:i8[N,512], B:i8[512,M], C:i8[N,M]):
  assert N % 256 == 0
  assert M % 256 == 0
  for i in seq(0, N):
    for j in seq(0, M):
      res: i32 @ DRAM
      res = 0.0
      for k in seq(0, 512):
        a2: i32 @ DRAM
        b2: i32 @ DRAM
        a2 = A[i, k]
        b2 = B[k, j]
        res += a2 * b2
      src_tmp: i32 @ DRAM
      src_tmp = res
      tmp_res1: f32 @ DRAM
      acc_scale(src_tmp, tmp_res1, scale)
      tmp_res2: i8 @ DRAM
      clamp(tmp_res1, tmp_res2)
      if act == True:
        tmp_res2 = relu(tmp_res2)
      C[i, j] = tmp_res2
```

***Hardware Constraints and Cursors.*** Gemmini has a software-managed scratchpad of size 256KB and an accumulator of size 16KB. To efficiently utilize these hardware resources, the memory layout of the scratchpad and accumulator must have innermost dimensions of 16. We obtain cursors to key points in the object code, which will be used throughout the scheduling process to navigate and transform the code.

```
# Parameters
accum_size= 16 * 1024  # 16 KB
sc_size   = 256 * 1024 # 256 KB
# Grab cursors
i         = p.find_loop("i")
j         = p.find_loop("j")
k         = p.find_loop("k")
res_load  = p.find("res = 0.0")
a_assign  = p.find("a2 = A[_]")
b_assign  = p.find("b2 = B[_]")
res_alloc = res_load.prev()
```

***Tiling for Scratchpad and Accumulator.*** To ensure that the working set fits into the scratchpad and accumulator, we tile the loops using the `tile_loops`, which is a user-level Gemmini library function. This function divides each loop into an outer and inner loop, rearranges the outer loops to be outside the inner loops, and returns cursors to the inner loops. In the provided schedule, we apply two levels of tiling to the `i` and `j` loops, while the `k` loop is simply divided by 16.

```
p, _ = tile_loops(p, [(i,16), (j,16)], perfect=True)
p, [_, j_outer] = tile_loops(p, [(i,16), (j,16)],
                                       perfect=True)
p, _ = tile_loops(p, [(k, 16)])
```

***Memory Allocation and Lifting.*** `autolift_alloc` and `bind_and_lift` allocate buffers in the scratchpad and accumulator. `autolift_alloc` lifts and expands buffer dimensions within the specified threshold (accumulator and scratchpad sizes). `bind_and_lift` creates and lifts a temporary buffer using `autolift_alloc`. We allocate an `i32[16, 16, 16]` buffer in the accumulator, matching its size. For input matrices `A` and `B`, we bind them to temporary scratchpad memory and lift the allocations. The resulting `A_tmp` and `B_tmp` buffers each occupy 128KB, half of the scratchpad.

```
p = autolift_alloc(p, res_alloc, max_size=accum_size)
p, a_load, a_alloc = bind_and_lift(p,
        p.forward(a_assign).rhs(), max_size=sc_size)
p, b_load, b_alloc = bind_and_lift(p,
        p.forward(b_assign).rhs(), max_size=sc_size)
```

***Utilizing Blocked Load Instructions.*** Although Gemmini's matrix multiplication instruction operates on $16 \times 16$ tiles, it supports larger blocked loads of size $4 \times 16 \times 16$ in a single cycle. To take advantage of this feature, we further divide the `k` and `j` loops by 4.

```
p, _ = tile_loops(p, [(k_loop, 4)])
p, [j_imost] = tile_loops(p, [(j_outer, 4)],
                                  perfect=True)
```

***Loop Fission and Instruction Mapping.*** We apply loop fission to each code block, isolating individual chunks that correspond to Gemmini instructions.

```
p = fission_as_much_as_possible(p, res_load)
p = fission_as_much_as_possible(p, k)
p = fission_as_much_as_possible(p, a_load)
p = fission_as_much_as_possible(p, b_load)
```

***Data Layout Transformation.*** The dimensions of `A_tmp` and `B_tmp` buffers need to be reordered to match the expected layout of Gemmini instructions. Specifically, we use the `rearrange_dim` function to ensure that the innermost two dimensions of `A_tmp` are along the `i` and `k` axes, while dimensions for `B_tmp` are along the `k` and `j` axes.

```
p = rearrange_dim(p, a_alloc, [0, 2, 1, 3])
p = rearrange_dim(p, b_alloc, [2, 0, 3, 1])
p = reorder_loops_from_idx(p, a_load)
p = reorder_loops_from_idx(p, b_load)
p = reorder_loops_from_idx(p,
                p.forward(a_assign).rhs())
p = remove_redundant_loops(p, a_load, num=2)
p = remove_redundant_loops(p, b_load, num=1)
```

***Replacing with Gemmini Instructions and Hoisting Configuration.*** Gemmini instructions require setting certain configuration states. In Exo, the instructions are represented by pairs of functionally equivalent procs, with one version having additional configuration state updates. For instance, `ld_i8_block_id1_v2` writes to a configuration stride register before calling `ld_i8_block_id1`. We use the `replace_and_inline` function to first insert calls to the `_v2` instructions, and then inline to replace it with a configuration update and the original instructions. Then, the configuration instructions are hoisted out of the loops to minimize their overhead (see Section 6.1 for user-level scheduling functions that support configuration hoisting).

```
p = set_memory(p, res_alloc, GEMM_ACCUM)
p = set_memory(p, a_alloc, GEMM_SCRATCH)
p = set_memory(p, b_alloc, GEMM_SCRATCH)
tuples = [
    (zero_acc_i32, zero_acc_i32_v2),
    (ld_i8_block_id1, ld_i8_block_id1_v2),
    (ld_i8_block_id2, ld_i8_block_id2_v2),
    (matmul_acc_i8, matmul_acc_i8_v2),
    (st_acc_i8, st_acc_i8_v2)]
for t in tuples:
    p = simplify(replace_and_inline(p, t))
```

***Optimizing Memory Access.*** We add guards to avoid issuing redundant load instructions while keeping the loop structure intact for further optimization.

```
p = add_guard(p, p.find("do_ld_i8_block_id1(_)"))
p = add_guard(p, p.find("do_ld_i8_block_id2(_)"))
```

***Loop Fusion and Unrolling.*** To prevent the reorder buffer (ROB) from being filled with only load instructions, we fuse the loops to achieve a balanced mix of load, store, and compute instructions. This improves locality and exploits instruction-level parallelism. We also unroll all the innermost loops to reduce loop overhead.

```
p = fuse_all_loops(p, p.body()[0])
p = unroll_all(p, p.find_loop(j_imost.name(),
                         many=True))
```

***Final object code.*** The final object code, just before the unrolling step, is shown on the right. This function takes arbitrary N and M as long as they are multiples of 256 (lines 3-4). As we showed in Section 6.1.2, configuration instructions are hoisted at the beginning of the kernel, eliminating the overhead of reconfiguration (lines 5-9). Buffers for A, B, and C are allocated in the Gemmini accumulator and scratchpad memories (lines 10-12). In the main loop, it first initializes the accumulator to zero (lines 17-19), loads A and B to the scratchpad (lines 21-33), and performs matrix multiplication on the accumulator (lines 34-38). Finally, it stores the result from the accumulator to C (lines 39-44). This optimized code efficiently utilizes Gemmini's scratchpad and accumulator memories and leverages its instructions and configuration states to achieve high-performance matmul.

```
def matmul_on_gemmini(N: size, M: size, scale: f32,
    act:bool, A:i8[N,512], B:i8[512,M], C:i8[N,M]):
  assert N % 256 == 0
  assert M % 256 == 0
  config_st_acc_i8(scale, stride(C, 0), act)
  config_matmul()
  config_ld_i8_id2(stride(B, 0))
  config_ld_i8_id1(stride(A, 0))
  config_zero()
  res: i32[16, 16, 16] @ GEMM_ACCUM
  A_tmp: i8[16, 32, 16, 16] @ GEMM_SCRATCH
  B_tmp: i8[32, 16, 16, 16] @ GEMM_SCRATCH
  for ioo in seq(0, N / 256):
    for joo in seq(0, M / 256):
      for ioi in seq(0, 16):
        for joio in seq(0, 4):
          for joii in seq(0, 4):
            do_zero_acc_i32(16, 16,
                       res[joii+4*joio, 0:16, 0:16])
          for koo in seq(0, 8):
            if joo == 0:
              if joio == 0:
                do_ld_i8_block_id1(16, 4,
                 A[16*ioi+256*ioo:16+16*ioi+256*ioo,
                                  64*koo:64+64*koo],
                 A_tmp[ioi,4*koo:4+4*koo,0:16,0:16])
            for koi in seq(0, 4):
              if ioi == 0:
                do_ld_i8_block_id2(16, 4,
                  B[16*koi+64*koo:16+16*koi+64*koo,
                64*joio+256*joo:64+64*joio+256*joo],
                  B_tmp[koi+4*koo, 4*joio:4+4*joio,
                                        0:16, 0:16])
              for joii in seq(0, 4):
                do_matmul_acc_i8(16, 16, 16,
                  A_tmp[ioi, koi+4*koo, 0:16, 0:16],
                 B_tmp[koi+4*koo, joii+4*joio, 0:16,
                0:16], res[joii+4*joio, 0:16, 0:16])
          for joii in seq(0, 4):
            do_st_acc_i8(16, 16,
                  res[joii+4*joio, 0:16,0:16],
                 C[16*ioi+256*ioo:16+16*ioi+256*ioo,
                    16*joii+64*joio+256*joo:
                    16+16*joii+64*joio+256*joo])
```

## C Matrix-multiply on AVX512

***Initial Object Code.*** This section provides a step-by-step guide for optimizing matrix multiplication on AVX512. Our starting point is the simple triple nested-loop matrix multiplication implemented as an outer product. This section does not include every code snippet evaluated in Section 6.2.3. Specifically, it omits the code for the bottom and right panels. However, these panels follow a strategy similar to that of the main panel.

```
@proc
def SGEMM(M: size, N: size, K: size,
          A: f32[M, K], B: f32[K, N],
          C: f32[M, N]):
    for k in seq(0, K):
        for i in seq(0, M):
            for j in seq(0, N):
                C[i, j] += A[i, k] * B[k, j]
```

***Hardware Constraints.*** Before diving into the optimization process, we define hardware constraints such as the vector width (`VEC_W`), register blocking factors (`M_REG_BLK`, `N_REG_BLK`), and L1 cache tiling factors (`M_L1_FAC`, `N_L1_FAC`, `K_L1_BLK`).

Our optimization strategy consists of two main stages. First, we tile the loops to optimize for the memory hierarchy, improving data locality and reducing cache misses. Second, we compute the matrix-multiplication of these tiles by performing register-tiling.

```
VEC_W = 16
M_REG_BLK = 6
N_REG_BLK = 4 * VEC_W
M_L1_FAC = 44
N_L1_FAC = 1
M_L1_BLK = M_REG_BLK * M_L1_FAC
N_L1_BLK = N_REG_BLK * N_L1_FAC
K_L1_BLK = 512
```

***Memory System Blocking and Staging.*** Memory-level tiling decomposes the overall multiplication into `M_L1_BKL x N_L1_BLK x K_L1_BLK` sub-problems. We start from the triple-nested outer product matrix multiplication, and we tile it to generate the main tile case and the various tail cases. Then, we stage the `B` tile into local static memory for each of the cases. We only stage the `A` tile for the main tile. We then hoist the loop nest of the load stage of `A` since it is idempotent with respect to the innermost loop. Finally, we replace all the main tile case and all the tail cases with the scheduled matrix-multiplication that calls to the various microkernels that are scheduled in the following paragraphs.

```
p = rename(SGEMM, "sgemm_exo")
p = tile_loops_bottom_up(p, p.body()[0],
                         (K_L1_BLK, M_L1_BLK, N_L1_BLK))
for i in range(0, 8):
  B_name = "B_cache"
  p, (alloc, _, _, _) = auto_stage_mem(p,
      p.find_loop(f"ki #{i}"), "B", B_name, rc=True)
  p = resize_dim(p, alloc, 0, K_L1_BLK, 0)
  p = resize_dim(p, alloc, 1, N_L1_BLK, 0)
  p = set_memory(p, alloc, DRAM_STATIC)
p, (alloc, _, _, _) = auto_stage_mem(p,
        p.find_loop("jo"), "A", "A_cache", rc=True)
p = resize_dim(p, alloc, 0, M_L1_BLK, 0)
p = resize_dim(p, alloc, 1, K_L1_BLK, 0)
p = set_memory(p, alloc, DRAM_STATIC)
p = fission(p, p.find_loop("jo").after(), n_lifts=2)
p = replace_all(p, SGEMM_WINDOW)
p = repeat(call_eqv)(p, SGEMM_WINDOW,
                                sgemm_above_kernel)
```

***Micro-Kernel Generation Function.*** Register-level tiling (micro-kernel) computes `N_REG_BKL x M_REG_BLOCK` of `C` as an outer product of micropanels from matrices `A` and `B`. Our primary micro-kernel computes a $6 \times 64$ microtile of $C$. For tail cases, we generate additional micro-kernels of size $m \times 16n$ where either $m = 6$ and $n \leq 4$, or $m \leq 6$ and $n = 4$. Microtiles with the second dimension with not multiples of 16 are rounded up and dispatched to the nearest appropriate micro-kernel.

A single, parameterized function `gen_ukernel` generates schedules for all micro-kernels. The parameter `cond` represents the multiple of 16 to which $n$ rounds up. The schedule first asserts conditions about `cond`, rounds the $j$ loop to a multiple of 16, stages the `C` microtile, and vectorizes the loops for loading, outer product computation, and storing. After applying dead code elimination to remove unnecessary branches, it replaces SIMD loops with AVX-512 instructions.

```
def gen_ukernel(p, cond):
  og_p = p.add_assertion(cond)
  p = round_loop(og_p, og_p.find_loop("j"), VEC_W)
  p = simplify(specialize(p, p.find_loop("k"), cond))
  p = dce(p)

  p = simplify(auto_stage_mem(p, p.find_loop("k"),
                                          "C", "C_reg"))
  p = simplify(divide_dim(p, "C_reg", 1, VEC_W))
  p = set_memory(p, "C_reg", AVX512)

  rules = [fma_rule]
  for it in ("i1", "j", "i1"):
    p = vectorize(
      p, p.find_loop(it), VEC_W, "f32", AVX512,
                       rules=rules, tail="perfect")
  p = hoist_from_loop(p, p.find_loop("jo"))
  for it in ("i1o", "jo", "i1o"):
    p = unroll_loop(p, it)

  p = simplify(dce(p))
  p = replace_all(p, AVX512F_instructions)
  return og_p, simplify(p)
```

***Generating Micro-Kernels.*** We use the `gen_ukernel` function to create all necessary micro-kernels. This is done by calling the function with procedures that are partially evaluated for different values of $M$. For the main and bottom micro-kernels, which don't require predicates, we set `cond` to True (`"0 == 0"`). When generating right panels, we use an actual condition for `cond`.

The resulting micro-kernels are stored in the `basic_kernel_Mx4` and `sgemm_kernel_avx512_Mx4` dictionaries for later use.

```
basic_kernel_Mx4 = {}
sgemm_kernel_avx512_Mx4 = {}
for M in range(1, M_REG_BLK + 1):
  def make_basic(p):
    p = rename(p, f"basic_kernel_{M}x4")
    p = p.partial_eval(M, N_REG_BLK)
    p = simplify(p)
    return p

  p = make_basic(SGEMM_WINDOW)
  p, p_sched = gen_ukernel(p, "0 == 0")
  basic_kernel_Mx4[M] = p
  sgemm_kernel_avx512_Mx4[M] = p_sched
```

***Micro-Kernel Object Code.*** The code on the right shows an example of a generated micro-kernel for a 6 × 64 micro-tile. The kernel assumes that the input matrices A and B, as well as the output matrix C, are stored in row-major order with unit stride (lines 3-6). The micro-tile of C is staged into a local register array `C_reg` (line 7), which is then used for loading from C (lines 8-12), accumulating the results of the outer product between micropanels of A and B (lines 13-32), and storing the result back to C (lines 33-37). The computation is fully vectorized using AVX512 intrinsics, with each iteration of the innermost loop processing a 16-element vector. By applying the aforementioned optimizations, the matrix multiplication code is efficiently mapped to the AVX512 instruction set, taking full advantage of the available vector registers and SIMD parallelism.

```
def basic_kernel_6x4(K: size, A: [f32][6, K] @ DRAM,
     B: [f32][K, 64] @ DRAM, C: [f32][6, 64] @ DRAM):
  assert K >= 1
  assert stride(A, 1) == 1
  assert stride(B, 1) == 1
  assert stride(C, 1) == 1
  C_reg: f32[6, 4, 16] @ AVX512
  for i0 in seq(0, 6):
    mm512_loadu_ps(C_reg[i0, 0, 0:16], C[i0, 0:16])
    mm512_loadu_ps(C_reg[i0, 1, 0:16], C[i0, 16:32])
    mm512_loadu_ps(C_reg[i0, 2, 0:16], C[i0, 32:48])
    mm512_loadu_ps(C_reg[i0, 3, 0:16], C[i0, 48:64])
  for k in seq(0, K):
    for i in seq(0, 6):
      var0: f32[16] @ AVX512
      mm512_set1_ps(var0[0:16], A[i, k:1 + k])
      var1: f32[16] @ AVX512
      mm512_loadu_ps(var1[0:16], B[k, 0:16])
      mm512_fmadd_ps(var0[0:16], var1[0:16],
                              C_reg[i, 0, 0:16])
      var1_1: f32[16] @ AVX512
      mm512_loadu_ps(var1_1[0:16], B[k, 16:32])
      mm512_fmadd_ps(var0[0:16], var1_1[0:16],
                              C_reg[i, 1, 0:16])
      var1_2: f32[16] @ AVX512
      mm512_loadu_ps(var1_2[0:16], B[k, 32:48])
      mm512_fmadd_ps(var0[0:16], var1_2[0:16],
                              C_reg[i, 2, 0:16])
      var1_3: f32[16] @ AVX512
      mm512_loadu_ps(var1_3[0:16], B[k, 48:64])
      mm512_fmadd_ps(var0[0:16], var1_3[0:16],
                              C_reg[i, 3, 0:16])
  for i0 in seq(0, 6):
    mm512_storeu_ps(C[i0, 0:16], C_reg[i0, 0, 0:16])
    mm512_storeu_ps(C[i0, 16:32],C_reg[i0, 1, 0:16])
    mm512_storeu_ps(C[i0, 32:48],C_reg[i0, 2, 0:16])
    mm512_storeu_ps(C[i0, 48:64],C_reg[i0, 3, 0:16])
```

# D BLAS

## D.1 Level 1 Specific Scheduling Operator

```
def optimize_level_1(proc, loop, precision, machine, interleave_factor,
                                            vec_tail=None, inter_tail="recursive"):
    vec_width = machine.vec_width(precision)
    instrs = machine.get_instructions(precision)
    memory = machine.mem_type
    patterns = machine.patterns

    if vec_tail is None:
        vec_tail = "cut_and_pred" if machine.supports_predication else "cut"

    loop = proc.forward(loop)
    proc = CSE(proc, loop.body(), precision)

    proc, (loop,) = vectorize(proc, loop, vec_width, memory, patterns, vec_tail)

    proc, (_, loop) = LICM(proc, loop, rc=True)

    proc = interleave_loop(proc, loop, interleave_factor, memory, inter_tail)

    proc = cleanup(proc)
    proc = replace_all_stmts(proc, instrs)
    return proc
```

`optimize_level_1` is a scheduling operator we defined to optimize the entire BLAS level 1, achieving competitive performance against existing libraries across the board (Figures 14, 15, and 16). This operator optimizes a loop for a target machine at a given precision, taking arguments such as the interleaving factor and tail strategies. The implementation closely mirrors the body of the paper as discussed in Section 6.2.1. First, it extracts machine information, including vector width, instructions, and memory type (lines 3-6). For simplicity, CSE is performed before vectorization (line 12), and the `vectorize` function autovectorizes the loop as described in Section 6.1.1 (line 14). LICM then hoists broadcast instructions after vectorization (line 16). Loop iterations may be interleaved (line 18) to exploit SIMD vector parallelism. Finally, `replace_all_stmts` is called to replace the code with machine-specific vector instructions.

## D.2 Level 2 Specific Scheduling Operator

```
def optimize_level_2_general(proc, o_loop, precision, machine, r_fac, c_fac, round_up=None):
    o_loop = proc.forward(o_loop)

    proc, (o_loop,) = adjust_triang(proc, o_loop, machine, r_fac, round_up)

    proc = unroll_and_jam(proc, o_loop, r_fac)
    inner_loop = get_inner_loop(proc, o_loop)
    proc = optimize_level_1(proc, inner_loop, precision, machine, c_fac)

    tail = get_inner_loop(proc, proc.forward(o_loop).next())
    proc = optimize_level_1(proc, tail, precision, machine, c_fac)
    return cleanup(proc)
```

`optimize_level_2_general` optimizes 50 kernel variants, supporting most BLAS level-2 operations (excluding banded operations and packed-triangular formats) across all configurations. The performance is on par with existing libraries, as shown in Figures 17, 18, and 19. The function takes the outer loop (`o_loop`) of the level 2 kernel as input, along with the precision at which the kernel is computed and the target machine. Additionally, users can specify the number of rows (`r_fac`) to be batched and the number of columns (`c_col`) to be interleaved. A `round_up` parameter is provided to allow users to override the rounding behavior in triangular matrices. The implementation is simple and closely follows the body of the paper, as discussed in Section 6.2.2. First, the iteration space is adjusted for triangular matrices (line 4). Then, unroll-and-jam is performed to transform the loop into a batched program of multiple dot products (line 6). Finally, `optimize_level_1` is called on the resulting inner loop (line 11). More details about the Exo 2 BLAS library implementation can be found in [19].

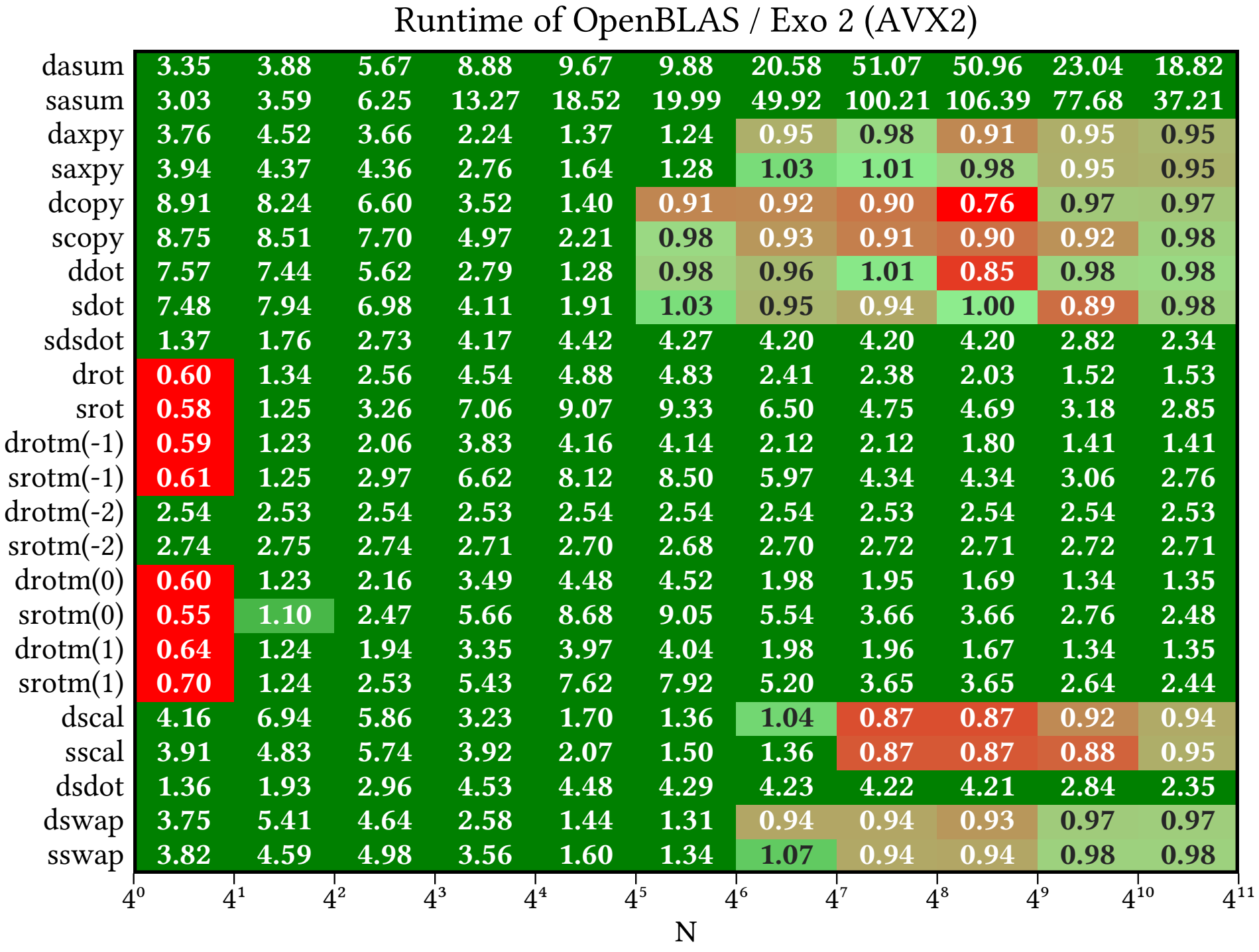

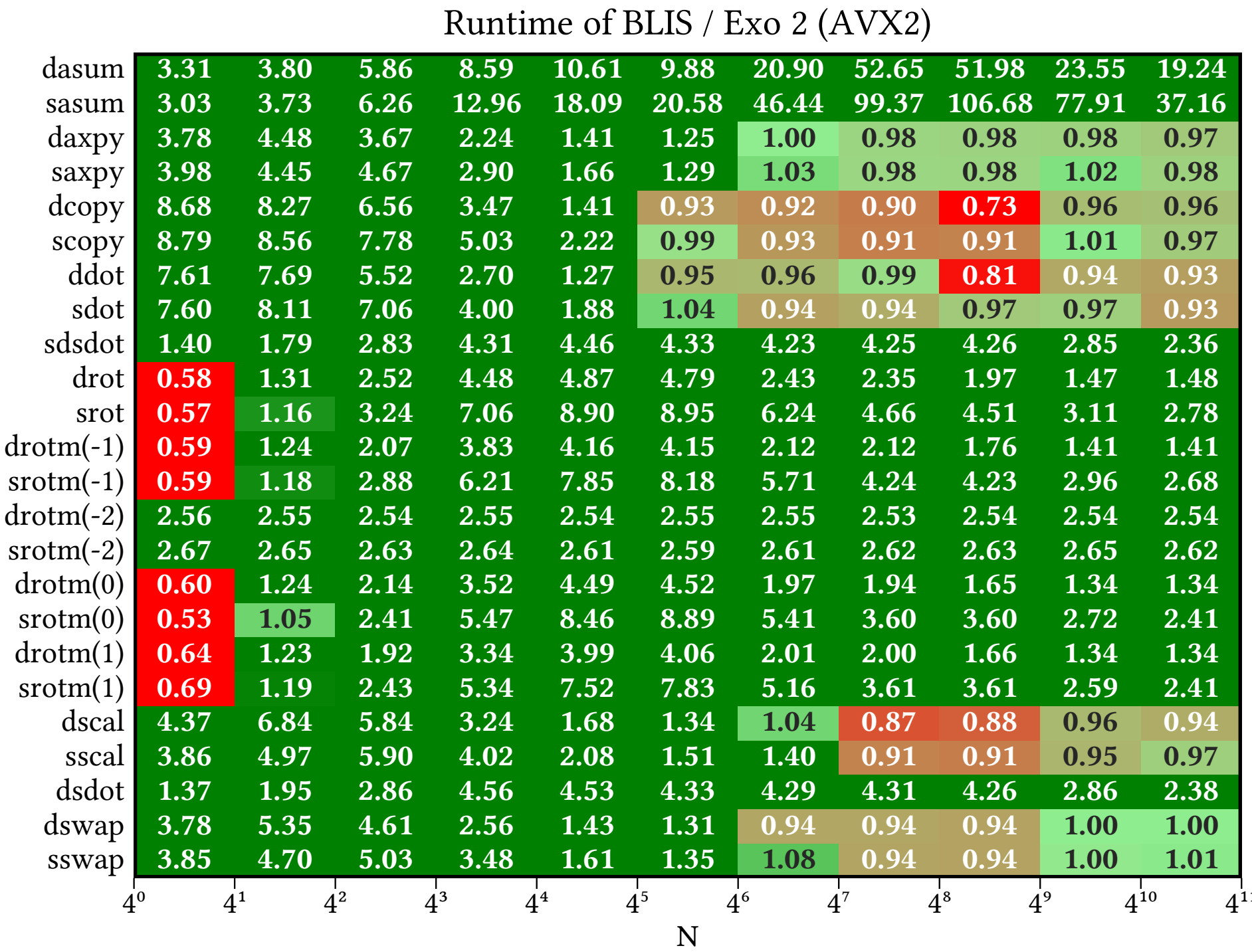

**Figure 14.** Exo 2 BLAS level 1 performance compared to OpenBLAS (top) and BLIS (bottom) using AVX2. Each heatmap cell represents the geometric mean over problems fitting into the respective x-axis bucket. Higher is better.

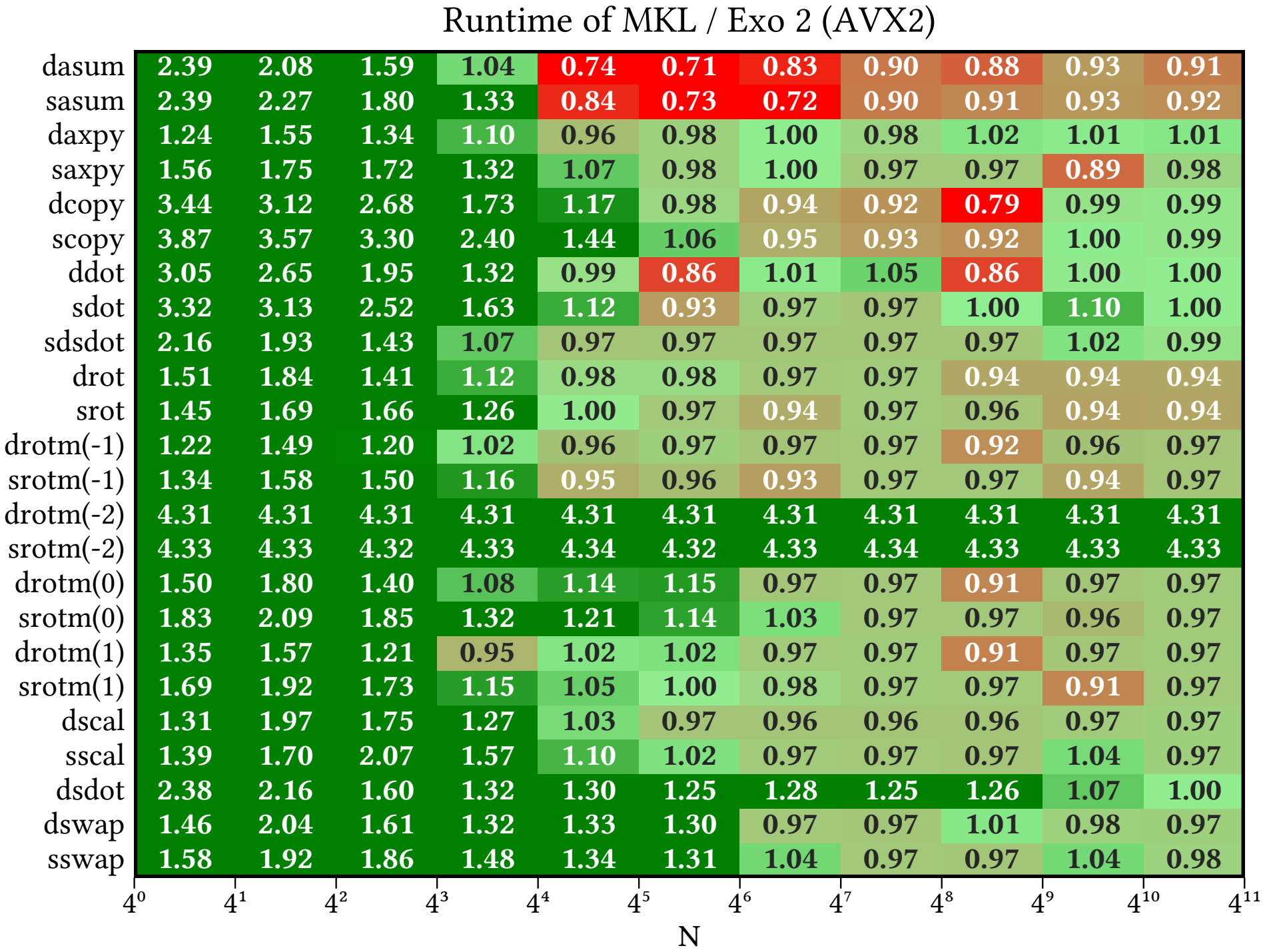

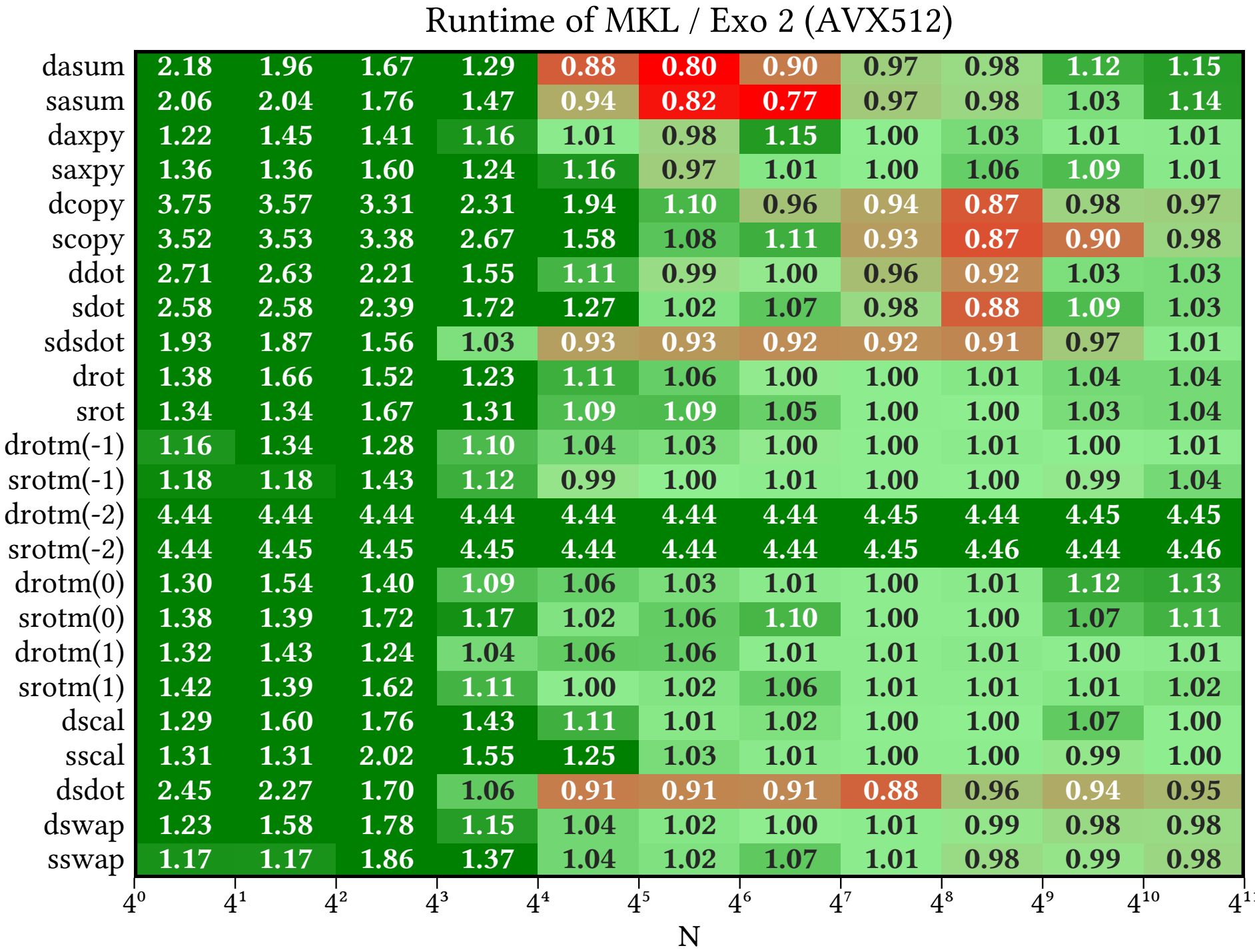

**Figure 15.** Exo 2 BLAS level 1 performance compared to MKL using AVX2 (top) and AVX512 (bottom).

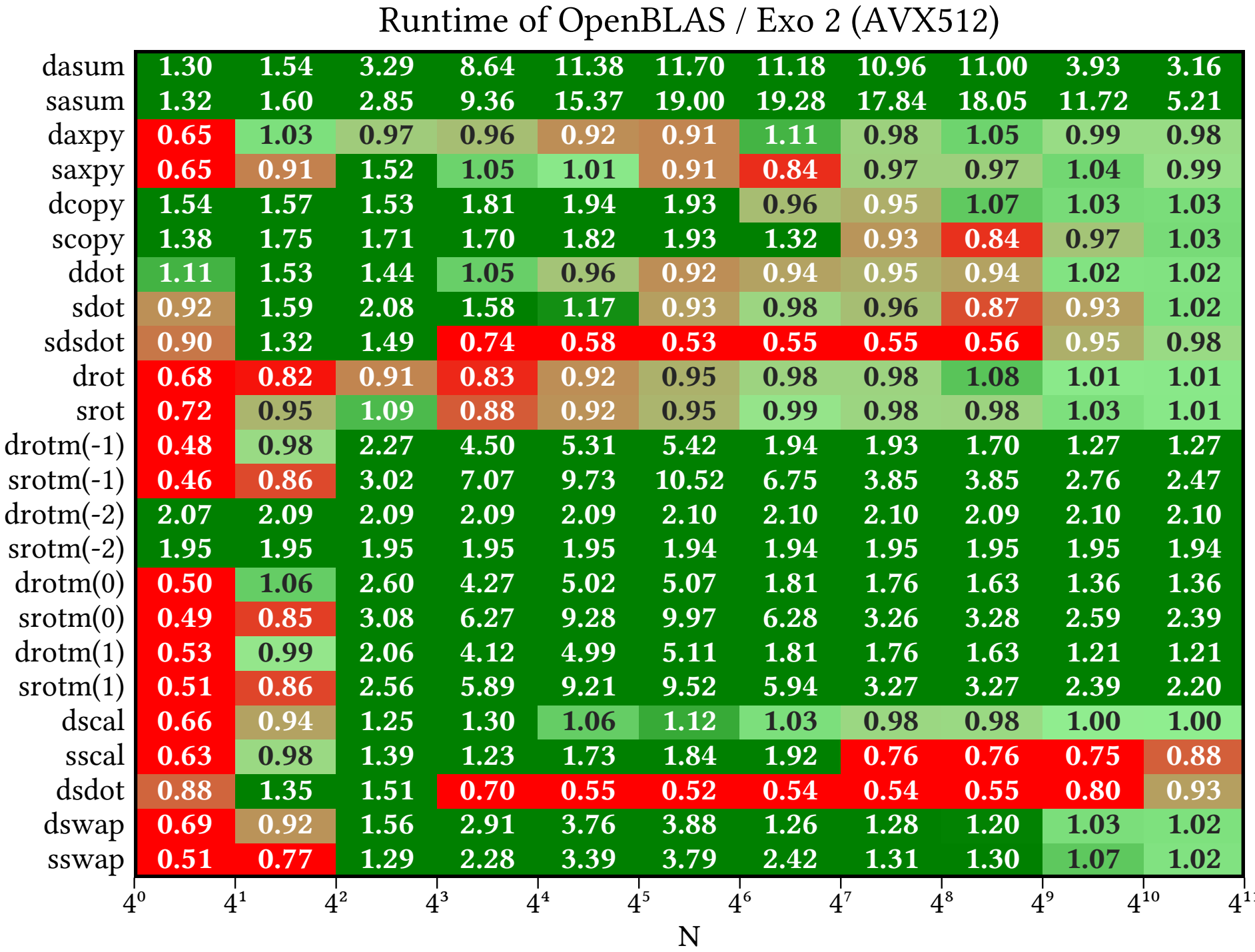

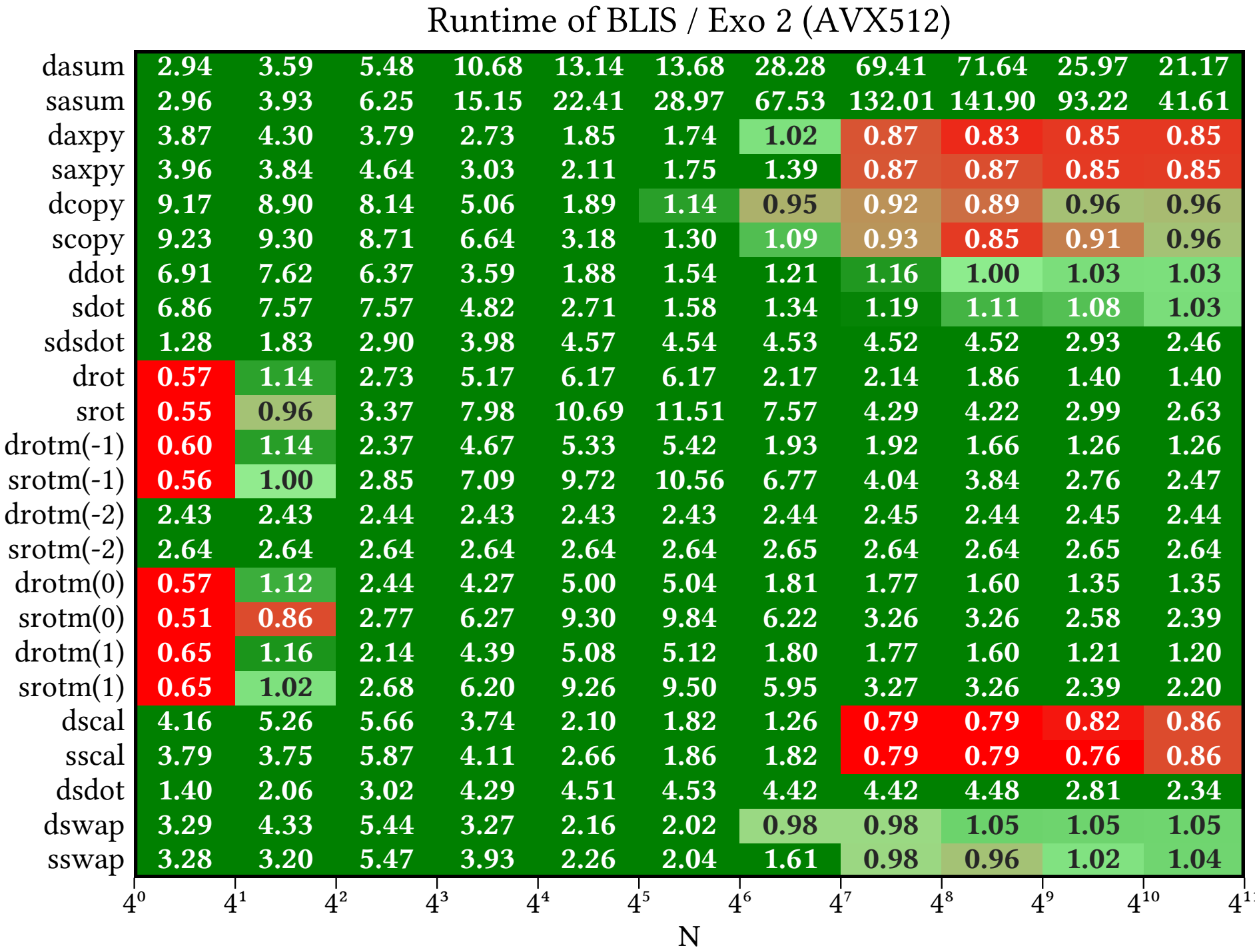

**Figure 16.** Exo 2 BLAS level 1 performance compared to OpenBLAS (top) and BLIS (bottom) using AVX512.

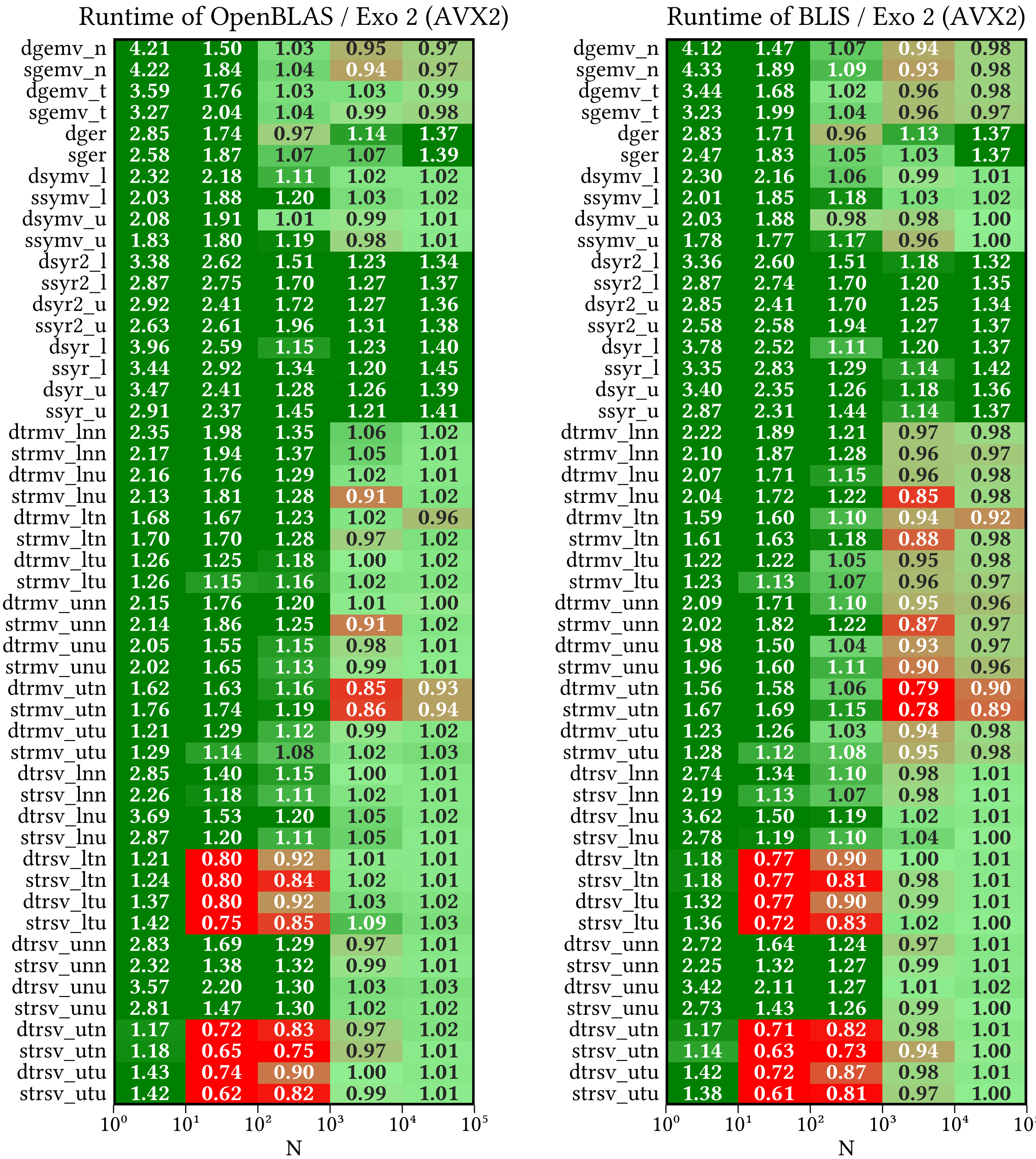

**Figure 17.** Exo 2 BLAS level 2 performance compared to OpenBLAS (left) and BLIS (right) using AVX2.

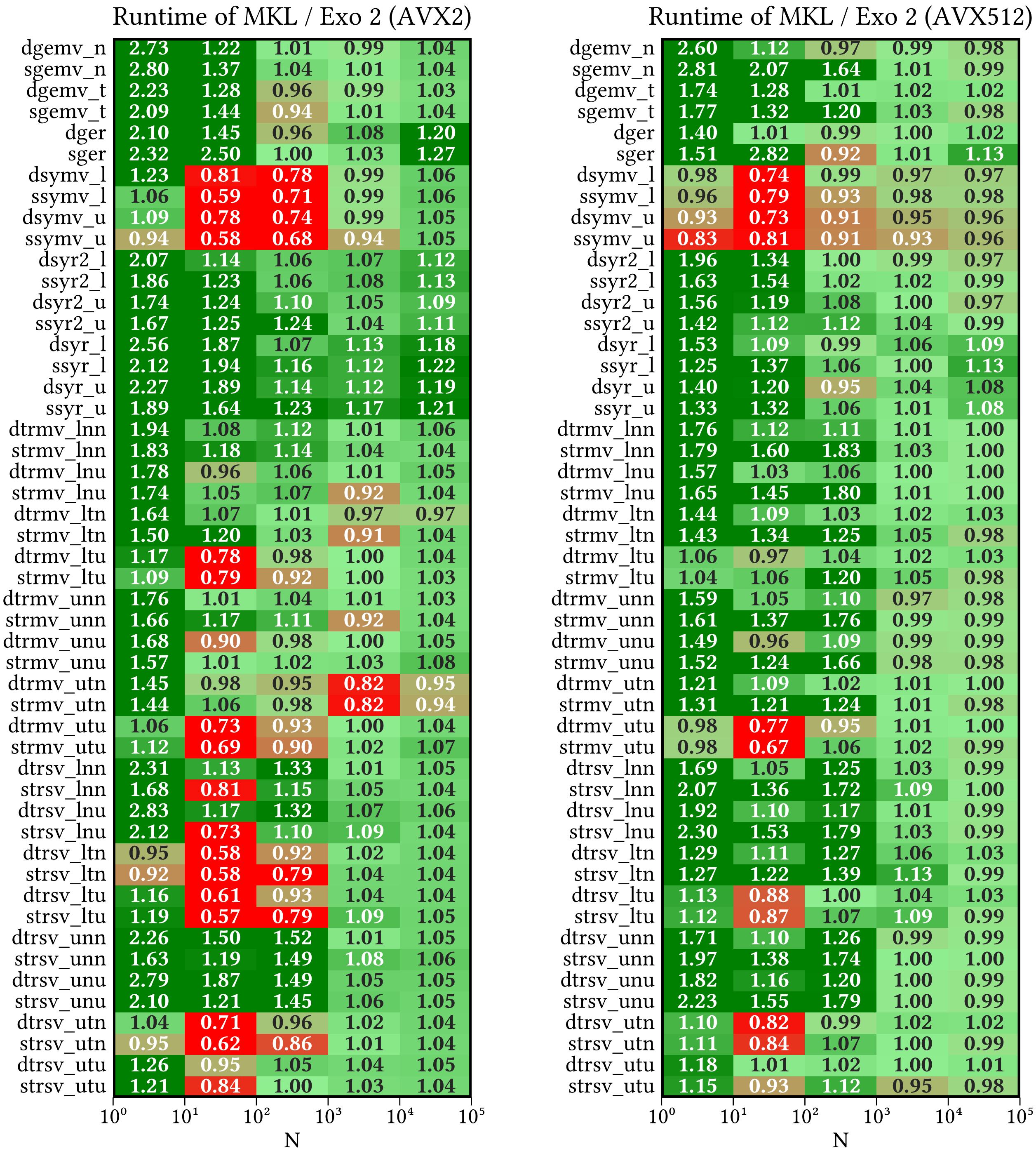

**Figure 18.** Exo 2 BLAS level 2 performance compared to MKL using AVX2 (left) and AVX512 (right).

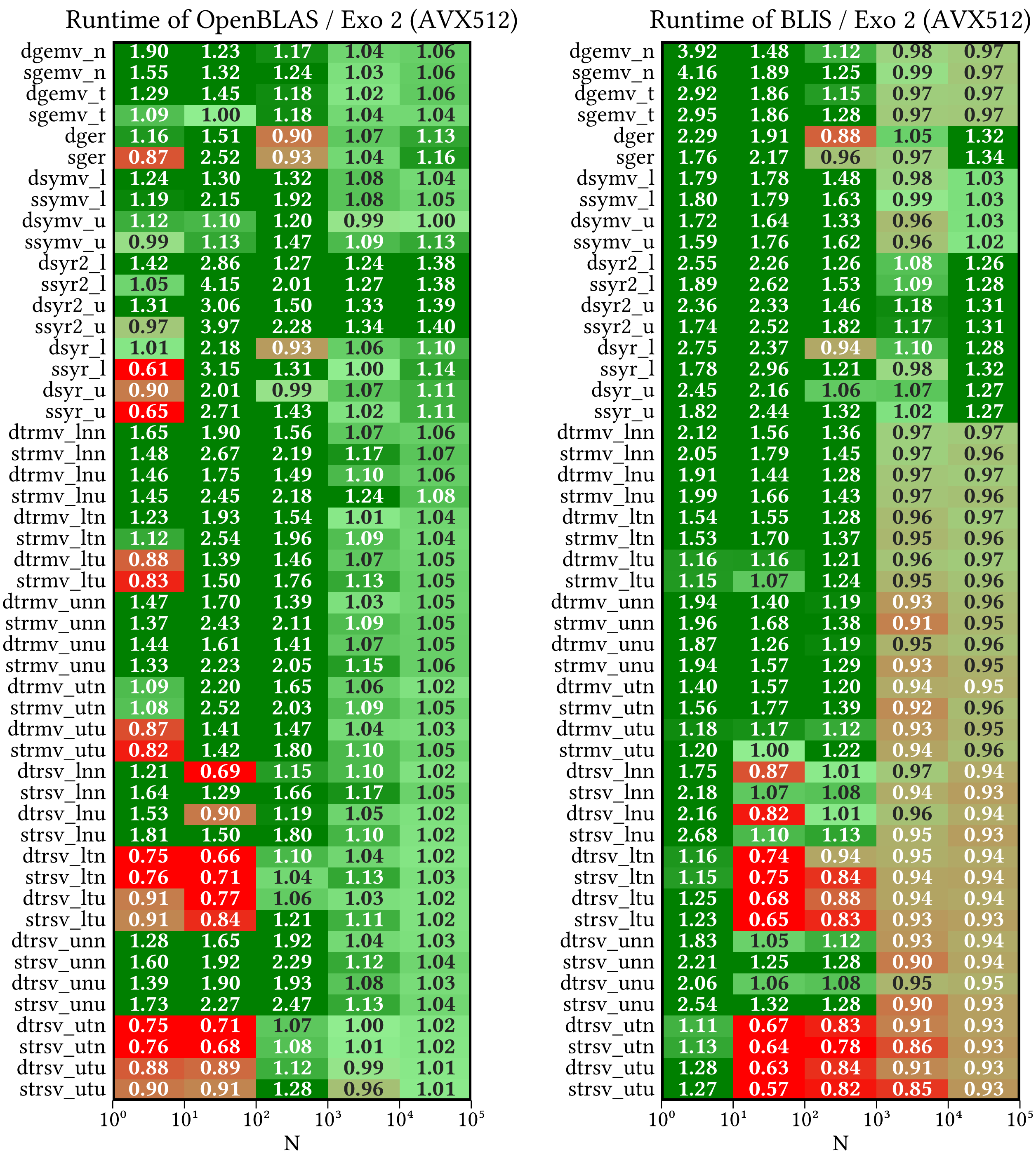

**Figure 19.** Exo 2 BLAS level 2 performance compared to OpenBLAS (left) and BLIS (right) using AVX512.